\documentclass[twocolumn,superscriptaddress,showpacs,prd,aps,amsmath,amssymb,nofootinbib]{revtex4}
\usepackage{graphicx}
\usepackage{color}
\usepackage{grffile}
\usepackage{amssymb}
\usepackage{bm} 
\usepackage{mathrsfs} 
\usepackage[normalem]{ulem}

\newcommand{\Madm}{M_{\rm ADM}}
\newcommand{\MK}{M_{\rm K}}

\newcommand{\Msol}{M_\odot}

\newcommand{\beq}{\begin{equation}} 
\newcommand{\eeq}{\end{equation}} 
\newcommand{\beqn}{\begin{eqnarray}} 
\newcommand{\eeqn}{\end{eqnarray}} 
\newcommand{\pa}{\partial}

\newcommand{\tgmabd}{\tilde\gamma_{ab}}
\newcommand{\tgamma}{\tilde\gamma}

\newcommand{\habd}{h_{ab}}

\newcommand{\fabd}{f_{ab}}

\newcommand{\tbeta}{\tilde{\beta}}

\newcommand{\albe}{{\alpha\beta}}

\newcommand{\Fabu}{F^{\alpha\beta}}

\newcommand{\zD}{{\raise1.0ex\hbox{${}^{\ \circ}$}}\!\!\!\!\!D}
\newcommand{\alone}{{\raise0.5ex\hbox{${}^{\ 1}$}}\!\!\!\!\alpha}

\newcommand{\dl}{\delta}

\newcommand{\Dl}{\Delta}

\newcommand{\compa}{M/R}

\newcommand{\nalam}{\mathrel{\raise0.9ex\hbox{$^\lambda$}\mkern-14mu
\lower0.0ex\hbox{$\nabla$}}}

\newcommand{\Nrf}{{N_r^{\rm f}}}
\newcommand{\Nrm}{{N_r^{\rm m}}}

\newcommand{\zeroD}{{\raise1.0ex\hbox{${}^{\ \circ}$}}\!\!\!\!\!D}

\newcommand{\zLap}{{\raise1.0ex\hbox{${}^{\ \circ}$}}\!\!\!\!\Delta}
\newcommand{\zna}{{\raise1.0ex\hbox{${}^{\ \circ}$}}\!\!\!\!\!\nabla}
\newcommand{\zS}{{\raise1.0ex\hbox{${}^{\ \circ}$}}\!\!\!\!\!S}
\newcommand{\hu}{h{\underbar u}}

\newcommand{\cocal}{{\sc cocal}}

\newcommand{\Bpolmax}{{B^{\rm max}_{\rm pol}}}
\newcommand{\Btormax}{{B^{\rm max}_{\rm tor}}}

\newcommand{\Bpol}{{B_{\rm pol}}}
\newcommand{\Btor}{{B_{\rm tor}}}

\newcommand{\rhoc}{\rho_{\rm c}}
\newcommand{\prhoc}{(p/\rho)_{\rm c}}
\newcommand{\Omegac}{\Omega_{\rm c}}

\begin{document}
\title{Equilibriums of extremely magnetized compact stars with 
force-free magnetotunnels}

\author{K\=oji Ury\=u}
\email{uryu@sci.u-ryukyu.ac.jp}
\affiliation{
Department of Physics, University of the Ryukyus, Senbaru 1, 
Nishihara, Okinawa 903-0213, Japan}
\author{Shijun Yoshida}
\email{yoshida@astr.tohoku.ac.jp}
\affiliation{
Astronomical Institute, Tohoku University, Aramaki-Aoba, Aoba, Sendai 980-8578, Japan}
\author{Eric Gourgoulhon}
\email{eric.gourgoulhon@obspm.fr}
\affiliation{
Laboratoire Univers et Th\'eories, UMR 8102 du CNRS,
Observatoire de Paris, Universit\'e PSL, Universit\'e Paris Diderot, F-92190 Meudon, France}
\author{Charalampos Markakis} 
\email{c.markakis@qmul.ac.uk}
\affiliation{
DAMTP, Centre for Mathematical Sciences, University of Cambridge, Cambridge, CB3 0WA, UK}
\affiliation{National Center for Supercomputing Applications, 
University of Illinois at Urbana-Champaign, Urbana, IL 61801, USA}
\affiliation{
School of Mathematical Sciences, Queen Mary University of London, London, E1 4NS, UK}
\author{Kotaro Fujisawa}
\email{fujisawa@resceu.s.u-tokyo.ac.jp}
\affiliation{
Department of Physics, Graduate School of Science, University of Tokyo, 
Hongo 7-3-1, Bunkyo, Tokyo 113-0033, Japan}
\affiliation{
Department of Liberal Arts, Tokyo University of Technology, 
5-23-22 Kamata, Ota, Tokyo 144-8535, Japan}
\author{Antonios Tsokaros}
\email{tsokaros@illinois.edu}
\affiliation{
Department of Physics, University of Illinois at Urbana-Champaign, Urbana, IL 61801}
\affiliation{National Center for Supercomputing Applications, 
University of Illinois at Urbana-Champaign, Urbana, IL 61801, USA}
\author{Keisuke Taniguchi}
\email{ktngc@sci.u-ryukyu.ac.jp}
\affiliation{
Department of Physics, University of the Ryukyus, Senbaru 1, 
Nishihara, Okinawa 903-0213, Japan}
\author{Mina Zamani}
\email{m_zamani@znu.ac.ir}
\affiliation{
Department of Physics, University of Zanjan, 
P.O.~Box 45195-313, Zanjan, Iran}
%
%

\date{\today}  

\begin{abstract} 
We present numerical solutions for stationary and axisymmetric equilibriums 
of compact stars associated with extremely strong magnetic fields.  The interior 
of the compact stars is assumed to satisfy ideal magnetohydrodynamic (MHD) 
conditions, while in the region of negligible mass density the 
force-free conditions or electromagnetic vacuum are assumed.  Solving all 
components of Einstein's equations, Maxwell's equations, ideal MHD equations, 
and force-free conditions, equilibriums of rotating compact stars associated 
with mixed poloidal and toroidal magnetic fields are obtained.  It is found 
that in the extreme cases the strong mixed magnetic fields concentrating 
in a toroidal region near the equatorial surface expel the matter 
and form a force-free toroidal magnetotunnel.  We also introduce a new 
differential rotation law for computing solutions associated with force-free 
magnetosphere, and present other extreme models without the magnetotunnel.  
\end{abstract} 


\maketitle
 
\section{Introduction}

Models for magnetars or hyper massive 
remnants of binary neutron star mergers are considered to be 
associated with strong magnetic fields around $10^{14}$--$10^{15}$G 
at their surfaces \cite{Magnetar,Magnetar_review}.  
Such strong magnetic fields plays essential roles in generating 
the signals arriving from the Soft Gamma Repeater and anomalous X-ray pulsar, 
as well as the jet of the merger remnants.  These magnetic fields 
are stronger than any other observed objects, and their interior magnetic fields 
could be one or two orders of magnitude stronger.  It is, however, far 
too weak to modify the equilibrium structure of the compact stars.  
Therefore in a theoretical modeling of such compact stars, 
the electromagnetic fields may be treated as a perturbation, and hence 
separately from the hydrostatic equilibriums of the stars 
(see e.g. \cite{PolTorMS,Tsokaros:2021tsu}).  

Although it may be astrophysically unrealistic in the present 
universe, it is theoretically interesting to further investigate 
the compact stars associated with extremely strong magnetic fields -- 
strong enough to modify the configuration of equilibrium structure 
and to dominate as a source of gravity.  It is also interesting theoretically 
to study the strongest limit of magnetic fields of the compact star. 
Such solutions may be of use for initial data sets for simulations \cite{MRNSsimulations}
from which the behavior of magnetic fields of compact stars would be elucidated 
within a relatively short simulation time \cite{Tsokaros:2021pkh}.  

In our previous papers \cite{Uryu:2014tda,Uryu:2019ckz}, we have 
developed a numerical code for stationary and axisymmetric equilibriums 
of relativistic rotating stars associated with extremely strong electromagnetic 
fields based on our versatile code library, \cocal\ (Compact Object 
CALculator), for calculating equilibriums and quasi-equilibrium initial 
data \cite{cocal}.  In this code, a full set of Einstein-Maxwell 
equations accompanied with ideal MHD equations are solved under assumptions 
of stationarity and axisymmetry consistently for the first time.  
In the paper \cite{Uryu:2019ckz} (hereafter Paper I), 
we have obtained solutions with mixed poloidal and toroidal 
magnetic fields whose exteriors are electromagnetic vacuum.  
We demonstrated that in the extreme cases the matter in a toroidal 
region near the equatorial surface was partly expelled by the extremely 
strong magnetic fields.  Such solutions have been used as initial data 
for fully numerical relativity simulations in \cite{Tsokaros:2021pkh}.  

In this paper, we introduce two new extensions to our previous 
works \cite{Uryu:2014tda,Uryu:2019ckz}, 
one is the force-free magnetic fields, and the other is the 
differential rotation.  In paper I, we assumed electromagnetic 
vacuum in the region where the energy density of the matter 
(or the rest mass density) became negligible following the 
idea of \cite{Bocquet:1995je}.  
The former extension is introduced to replace the electromagnetic 
vacuum with the force-free electromagnetic plasma region.  
Such low density regions appear not only at the exterior of the 
surface of a compact star, but also is expected to appear in the interior 
of it when the magnetic fields are stronger than the above mentioned 
solutions obtained in previous Paper I.  
To realize this, 
we follow a prescription for computing such compact stars associated 
with magnetosphere proposed by Southampton group \cite{Glampedakis:2013kaa}.  
This prescription has been also investigated by Florence group \cite{Pili:2014zna} 
for computing relativistic non-rotating stars under a simplified spatially 
conformal flat metric for the strong gravity (for numerical computations 
of magnetized relativistic stars, see also, \cite{Bocquet:1995je,MRNSequilibriums}).  
As we will see below, 
in the newly calculated solutions, it is found that the mixed poloidal 
and toroidal magnetic fields concentrate near, but well inside of, 
the equatorial surface, and that the fields totally expel the matter 
there, when the field strength becomes of the order of $10^{17}$G or 
higher for typical neutron stars, that is,  
a toroidal force-free magnetotunnel is formed in such a solution.  
We also calculated the solutions replacing the electromagnetic vacuum 
exterior with the force-free magnetosphere.  
The latter extension 
for introducing the differential rotation is used to compute a rotating 
magnetized compact stars surrounded by the magnetosphere.  For a comparison, 
differentially rotating models with electromagnetic vacuum exterior are 
also obtained.  

This paper is organized as follows.  In Sec.~\ref{sec:Formulation}, 
we formulate the force-free fields in terms of variables used in 
our previous works \cite{Uryu:2014tda,Uryu:2019ckz}, and then 
introduce details of the formulation adapted to the numerical method.  
In Sec.~\ref{sec:Results}, new numerical solutions combining 
the force-free fields, differential rotations, as well as the 
electromagnetic vacuum as in the previous paper are presented.

\section{Formulation and numerical method for 
the force-free electromagnetic fields}
\label{sec:Formulation}

In Paper I, we have detailed 
the formulation and numerical method for computing stationary and 
axisymmetric equilibriums of strongly magnetized relativistic rotating stars.  
A set of equations to be solved includes the rest mass conservation equation, 
ideal MHD conditions, MHD-Euler equations associated with 
the barotropic equation of state, and the magnetic and gravitational 
field equations.  From the consistency of the set of ideal MHD equations
under the stationarity and axisymmetry, a set of integrability conditions 
and first integrals are derived.  In particular, it requires an 
existence of a master potential $\Upsilon$ that several quantities 
(as well as combinations of quantities), for example, 
the $t$ and $\phi$ components of the electromagnetic 1-form 
$A_\alpha$ are the function of $\Upsilon$, 
$A_t = A_t(\Upsilon)$, $A_\phi = A_\phi(\Upsilon)$.

To obtain solutions in Paper I, we assumed that the star is 
ideal MHD fluid, and furthermore that the outside of the star 
is electromagnetic vacuum.  
Because the above dependences $A_t = A_t(\Upsilon)$ 
and $A_\phi = A_\phi(\Upsilon)$ are valid only on the support of 
ideal MHD fluid, the $A_t$ and $A_\phi$ 
components should be solved independently from Maxwell's equations 
in the electromagnetic vacuum region, although the latter assumption 
may not be astrophysically realistic.  

In this work, we introduce an option to replace the assumption 
for the electromagnetic vacuum with the force-free electromagnetic fields 
in the low density region out of the ideal MHD region.  
We introduce below the formulation for the force-free 
electromagnetic fields which is adapted to our formulation for 
magnetized compact stars in Paper I.  It turns out that only a 
minor modification is necessary to implement the force-free 
formulation into our previously developed code \cite{Uryu:2019ckz}.  

Hereafter, we use abstract index notation for tensors; 
the Greek letters $\alpha, \beta, \gamma, ...$ stand for 4D objects, 
the Latin lowercase letters $a, b, c, ...$ for spatial 3D objects, and 
the Latin uppercase letters $A, B, C, ...$ for meridional 2D objects.  
%
Also, we sometimes express the 2-forms, $F, dA$, and $d(\hu)$ 
omitting indices.  Such index-free notation may be used, with caution, 
for calculations involving forms and vectors.  
A dot denotes an inner product, that is, a contraction between adjacent 
indices. For example, a vector $v$ and a $p$-form $\omega$ have inner product 
\beq
v\cdot \omega = v^\gamma \omega_{\gamma\alpha\dots\beta}, \quad 
\omega\cdot v = \omega_{\alpha\dots\beta\gamma}v^\gamma .
\eeq

We introduce 3+1 form of the spacetime metric with 
conformally decomposed spatial metric, 
\beq
ds^2 = -\alpha^2 dt^2 +\psi^4 \tgmabd(dx^a+\beta^a dt)(dx^b+\beta^b dt)
\eeq
where $\alpha$, $\beta^a$, $\psi$, and $\tgmabd$ are the lapse, 
the shift, the conformal factor, and the conformally related spatial 
metric, respectively.  The $\tgmabd$ is 
further decomposed with the reference 3D flat metric $\fabd$ as 
$\tgmabd = \fabd + \habd$, and the conformal decomposition is constrained 
by a condition $\tgamma=f$ where $\tgamma$ and $f$ are the determinant of 
$\tgmabd$ and $\fabd$, respectively.  We further assume the spacetime is 
asymptotically flat and impose the Dirac gauge and the maximal slicing 
conditions as coordinate conditions.  

Further details in notations and a common part of formulations 
are found in Paper I.

\subsection{Force-free condition}
\label{sec:FF}

We assume stationarity and axisymmetry associated, respectively, 
with the timelike and spacelike Killing vectors $t^\alpha$ and 
$\phi^\alpha$.  These vectors are used as basis of vector and 
tensor quantities, for example, the current vector $j^\alpha$ 
may be written 
\beq
j^\alpha = j^t t^\alpha + j^\phi \phi^\alpha + j^A e_A^\alpha, 
\label{eq:current}
\eeq
where $e_A^\alpha$ are the coordinate basis of the other 
two spatial coordinates $x^A$, such as $r$ and $\theta$.

We assume that in the exterior of the ideal MHD fluid, 
the force-free electromagnetic field is carried by a certain plasma 
current whose density is negligible.  
Each component of the force-free conditions $F\cdot j = F_{\albe}j^\beta=0$
in stationary and axisymmetric system becomes as follows: 
\\ 
$t$-component:
\beqn
t\cdot(F\cdot j) 
&=& t\cdot(F\cdot e_A)j^A 
= j^A F_{tA} 
\nonumber\\
&=& -j^A\pa_A A_t=0, 
\label{eq:FF_t}
\eeqn
$\phi$-component:
\beqn
\phi\cdot(F\cdot j) 
&=& \phi\cdot(F\cdot e_A)j^A 
= j^A F_{\phi A} 
\nonumber\\
&=& -j^A\pa_A A_\phi=0, 
\label{eq:FF_phi}
\eeqn
$x^A$-components:
\beqn
&&
e_A\cdot(F\cdot j) 
= e_A\cdot(F\cdot t)j^t + e_A\cdot(F\cdot \phi)j^\phi
\nonumber\\
&&\qquad\qquad\quad\ 
+ e_A\cdot(F\cdot e_B)j^B
\nonumber\\
&&\qquad\quad
= F_{At}j^t + F_{A\phi}j^\phi + F_{AB}j^B
\nonumber\\
&&\qquad\quad
= j^t\pa_A A_t + j^\phi\pa_A A_\phi  + (dA)_{AB}j^B =0.
\ \ 
\label{eq:FF_xA}
\eeqn

The $x^A$ components of Maxwell's equations 
are written, 
\beq
4\pi j^A \sqrt{-g} \,=\, \pa_B(F^{AB}\sqrt{-g}) \,=\, \epsilon^{AB}\pa_B(\sqrt{-g}B), 
\label{eq:mercurrent}
\eeq 
where $B$ is defined as $F^{AB} = \epsilon^{AB} B$ 
(see, Sec.II.F.3 of Paper I).  

Substituting Eq.~(\ref{eq:mercurrent}) to $t$ and $\phi$ 
components of the force-free conditions 
Eqs.~(\ref{eq:FF_t}) and (\ref{eq:FF_phi}), 
we have, 
\beqn
&& 
\epsilon^{AB}\pa_A A_t \,\pa_B(\sqrt{-g}B) =0, 
\label{eq:idealMHD_t_form1}
\\[1mm]
&& 
\epsilon^{AB}\pa_A A_\phi \,\pa_B(\sqrt{-g}B) =0. 
\label{eq:idealMHD_phi_form1}
\eeqn
Hence, similarly to the case with ideal MHD fluid, 
the integrability conditions for the region 
of force-free electromagnetic fields can be written 
in terms of the master potential $\Upsilon$, 
\beqn
&&
A_t = A_t(\Upsilon), 
\quad\ \ 
A_\phi = A_\phi(\Upsilon),  
\nonumber\\[1mm]
&&\!\!\!\!\!\!\!\!\!\!\!\!\!\!\!\!\!\!
\mbox{and}
\qquad\quad
\sqrt{-g} B = [\,\sqrt{-g}B\,](\Upsilon).  
\label{eq:def_fn_At_Aphi_B}
\eeqn

Substituting the meridional component of the current ($x^A$ component 
of the Maxwell's equations) (\ref{eq:mercurrent}) and a 
definition $F_{AB} = \epsilon_{AB} B_\phi$, 
the $x^A$ component of the force-free condition (\ref{eq:FF_xA}) 
becomes
\beq
j^t\sqrt{-g}\pa_A A_t 
\,+\, j^\phi\sqrt{-g}\pa_A A_\phi
\,-\, \frac1{4\pi} B_{\phi}\pa_A[\sqrt{-g}B] \,=\,0.
\label{eq:FF_xA_form1}
\eeq
The conditions (\ref{eq:def_fn_At_Aphi_B}) imply 
\beq
\left(A'_t j^t\sqrt{-g}
\,+\, A'_\phi j^\phi\sqrt{-g}
\,-\,\frac1{4\pi}[\sqrt{-g}B]'B_{\phi}
\,\right)\pa_A \Upsilon
=0, 
\eeq
where the primes $A'_t$, $A'_\phi$ and $[\sqrt{-g}B]'$ 
stands for a derivative with respect to the master potential $\Upsilon$. 
Therefore, we have a consistency relation, which 
we also call first integral, for the stationary and axisymmetric 
force-free fields to satisfy, 
\beqn
A'_t j^t\sqrt{-g}
\,+\, A'_\phi j^\phi\sqrt{-g}
\,-\,\frac1{4\pi}[\sqrt{-g}B]'B_{\phi}
=0.  
\label{eq:fint_FF_xA}
\eeqn

The $t$ and $\phi$ components of Maxwell's equations 
are written, respectively
\beqn
4\pi j^t \sqrt{-g} &=& \pa_A(F^{t A}\sqrt{-g}),
\label{eq:current_t}
\\[1mm]
4\pi j^\phi \sqrt{-g} &=& \pa_A(F^{\phi A}\sqrt{-g}).
\label{eq:current_phi}
\eeqn 
Substituting either (\ref{eq:current_t}) or (\ref{eq:current_phi}) 
to Eq.~(\ref{eq:fint_FF_xA}), we have a relation for $j^\phi$ or $j^t$ 
to be used as a source term for an equation to determine the master 
potential, which is related to either $A_\phi(\Upsilon)$ or 
$A_t(\Upsilon)$, respectively.  So far in our actual numerical computations, 
we have been always choosing the 
master potential to be $\Upsilon = A_\phi$, and hence $\phi$ component 
of Maxwell's equations is used to determine the potential $A_\phi$ 
(see, Paper I).

\subsection{A model of Force-free field around an ideal MHD region}

In our formulation of ideal MHD presented in Paper I, we 
explicitly use the current $j^\alpha$ as an intermediate variable.  
Analogously, we have written down in the previous section \ref{sec:FF}
the force-free conditions as Eqs.(\ref{eq:def_fn_At_Aphi_B}) and 
(\ref{eq:fint_FF_xA}), whose forms are similar to the ideal MHD 
conditions, $F.u=0$.  
In actual computations, we choose $A_\phi$ as a master potential, 
$\Upsilon = A_\phi$.  Then, the integrability conditions and relations 
involves the current are written, 
\beqn
&&
A_t \,=\, A_t(A_\phi) 
\quad
\mbox{and}
\quad
\sqrt{-g} B \,=\, [\,\sqrt{-g}B\,](A_\phi),  
\quad
\label{eq:def_fn_At_B}
\\[3mm]
&&\qquad
j^A \sqrt{-g} 
\,=\, \frac1{4\pi}[\sqrt{-g}B]'\dl^{AB}B_B,
\label{eq:current_jA_FF}
\\[1mm]
&&\qquad
j^\phi\sqrt{-g}
\,+\, A'_t j^t\sqrt{-g}
\,=\,\frac1{4\pi}[\sqrt{-g}B]'B_{\phi} .  
\label{eq:current_jphi_FF}
\eeqn
where $B_A$ is defined by $\pa_A A_\phi=-\epsilon_A{}^B B_B$, 
$\delta^{AB}$ is the Kronecker delta.  

The corresponding expressions for the components of the current 
for ideal MHD fluid are (see, Eqs.~(135) and (136) of Paper I), 
\beqn
&&
j^A \sqrt{-g} 
\,=\,
\left([\sqrt{-g}\Psi]''hu_\phi
+[\sqrt{-g}\Lambda_\phi]'
\right)\delta^{AB}B_B
\nonumber\\
&&\qquad\qquad
\,-\,[\sqrt{-g}\Psi]'
\delta^{AB}\omega_B,  
\label{eq:current_jA_iMHD}
\\[2mm]
&&
j^\phi\sqrt{-g}
\,+\,A'_t\, j^t\sqrt{-g}
\nonumber\\
&&\quad
\,=\,
\left([\sqrt{-g}\Psi]''hu_\phi 
+ [\sqrt{-g}\Lambda_\phi]'\right)B_\phi
\,-\,[\sqrt{-g}\Psi]'\omega_\phi
\nonumber\\
&&\quad
\,-\,\left(A''_t \,hu_\phi+\Lambda'\right)\rho u^t \sqrt{-g}
\,-\, s' T \rho \sqrt{-g} , 
\label{eq:current_jphijt_iMHD}
\eeqn
where arbitrary functions $\sqrt{-g}\Lambda_\phi$ and $\Lambda$, 
the stream function $\sqrt{-g}\Psi$, and the entropy per 
baryon mass $s$ are functions of the master potential $A_\phi$
\beqn
&&
\sqrt{-g}\Lambda_\phi = [\sqrt{-g}\Lambda_\phi](A_\phi), 
\quad\ 
\Lambda = \Lambda(A_\phi), 
\nonumber\\[1mm]
&&\!\!
\sqrt{-g}\Psi = [\sqrt{-g}\Psi](A_\phi), 
\quad \mbox{and} \quad 
s = s(A_\phi).  
\label{eq:def_fn_Psi_Lp_L_s}
\eeqn
In Eqs.~(\ref{eq:current_jA_iMHD}) and (\ref{eq:current_jphijt_iMHD}), 
the terms including the stream functions $\sqrt{-g}\Psi$ 
are related to the meridional flow fields, $\rho$ is the rest mass 
density and $T$ the temperature.  
Since near the surface of compact stars, $\rho \rightarrow 0$, 
and $[\sqrt{-g}\Psi] \rightarrow \mbox{constant}$, terms related 
to the fluid approach to zero, and hence the remaining terms are 
\beqn
&&
j^A \sqrt{-g} 
\,\rightarrow\,
[\sqrt{-g}\Lambda_\phi]'\delta^{AB}B_B
\label{eq:current_jA_iMHD_limit}
\\[3mm]
&&
j^\phi\sqrt{-g}
\,+\,A'_t\, j^t\sqrt{-g}
\,\rightarrow\,
[\sqrt{-g}\Lambda_\phi]'B_\phi
\label{eq:current_jphi_iMHD_limit}
\eeqn
Therefore, comparing (\ref{eq:current_jA_FF}), (\ref{eq:current_jphi_FF}) 
and (\ref{eq:current_jA_iMHD_limit}), (\ref{eq:current_jphi_iMHD_limit}), 
we can smoothly connect the expressions of the current 
in the ideal MHD fluid region and in the force-free 
magnetosphere with the negligible density by choosing a common 
arbitrary function satisfying 
\beq
\frac1{4\pi} [\sqrt{-g}B]' = [\sqrt{-g}\Lambda_\phi]', 
\label{eq:iMHD_FF}
\eeq
in the whole domain of computation, 
and therefore connect the electromagnetic fields smoothly.

\subsection{Construction of magnetized star with 
magnetosphere/magnetotunnel}

As mentioned earlier, in the previous Paper I following \cite{Bocquet:1995je}, 
a region outside of the compact star was assumed to be the electromagnetic 
vacuum where the electric current vanishes.  Therefore, a component of 
vector potential $A_t$ was a function of $A_\phi$ on the ideal MHD 
fluid support, but was independent of $A_\phi$ otherwise.  
Assuming $A_\phi$ to be smooth across the stellar surface, we 
introduced (implicitly) the surface charge for $A_t$ to be 
continuous, but its normal derivative at the surface to be 
discontinuous.  Such a solution can be computed by adding 
a homogeneous function in solving $A_t$ to satisfy the above 
conditions at the surface.  

In our formulation for the force-free magnetosphere/magnetotunnel, 
we assume that the component $A_t$ is a function of $A_\phi$ in 
the whole domain and that a conducting current flows continuously 
and smoothly across the stellar surface.  Therefore, $A_t$ is no 
longer solved independently in the outside of the ideal MHD region.  
As it has been found in Paper I, when the toroidal magnetic 
field becomes extremely strong, the mixed poloidal and toroidal 
magnetic fields concentrate near the equatorial surface, and 
they expel the high density matter of compact stars.  It was, 
and, so far, it is not possible for the \cocal\ code to compute 
a toroidal vacuum tunnel inside of the compact star, because 
a method to impose a boundary condition to compute $A_t$ for 
such a toroidal region has not been developed yet.  On the 
other hand, if we assume a force-free field in such a toroidal 
region, the force-free magnetotunnel where the matter is expelled 
totally can be calculated straightforwardly 
under the above mentioned assumption in the same manner as 
computing the magnetosphere outside of the compact star.  
In latter sections, we will present such compact stars contain 
the magnetotunnel inside.

\subsubsection{Formulation for the fluid variables in equilibrium}

An equilibrium solution of the magnetized compact star can be 
calculated from a system of first integrals and integrability 
conditions derived in Paper I.  For choices with $\Upsilon = A_\phi$, 
and without meridional flows $[\sqrt{-g}\Psi](\Upsilon) = \mbox{constant}$, 
the following relations obtained from the normalization condition of 
4-velocity $u\cdot u=-1$, ideal MHD condition, and MHD-Euler equation 
(see, Paper I Sec.III.C), 

\beqn
u^t &=& \frac1{\left[\alpha^2-\psi^4\tgmabd
(v^a+\beta^a)(v^b+\beta^b)\right]^{1/2}}, 
\label{eq:ut}
\\[2mm]
\frac{u^\phi}{u^t}
&=& -A'_t \,=\, \Omega, 
\label{eq:uphi}
\\[2mm]
h \,
&=& \frac{\Lambda}{u_t - A'_t u_\phi} ,
\label{eq:h}
\eeqn
where the angular velocity of the matter $\Omega$ is also an arbitrary 
function of $A_\phi$ (Ferraro's law), 
\beq
\Omega = \Omega(A_\phi), 
\eeq
and the 4-velocity is written, 
\beq
u^\alpha 
\,=\, u^t (t^\alpha + v^\alpha)
\,=\, u^t (t^\alpha + \Omega\phi^\alpha),
\eeq
because the meridional components $u^A$ is assumed to vanish $u^A = u^t v^A =0$.

\subsubsection{Choice for arbitrary functions}

Further assuming the homentropic flow $s=\mbox{constant}$, 
five arbitrary functions of $A_\phi$, 
\beqn
&& \qquad 
A_t(A_\phi),\ \ \Omega(A_\phi),\ \ \Lambda(A_\phi), 
\nonumber \\[-2mm] \\[-2mm]
&& 
[\sqrt{-g}\Lambda_\phi ](A_\phi),\ \ \mbox{and}\ \ 
[\sqrt{-g}B](A_\phi), 
\nonumber
\label{eq:arbfnc}
\eeqn
appear in the above formulation for the matter (\ref{eq:ut})--(\ref{eq:h}) 
and the current (\ref{eq:current_jA_FF})--(\ref{eq:current_jphijt_iMHD}).  
Because of the assumption 
(\ref{eq:iMHD_FF}), four arbitrary functions remain to be specified.

For some arbitrary functions above, we choose, as in Paper I, 
a two parameter sigmoid function $\Xi'(x;b,c)$ 
that varies from $0$ to $1$ in between $0 < x < 1$ 
\beq
\Xi'(x;b,c) = \frac12\left[\,\tanh\left(\frac{x}{b}-c \right)+1\,\right],  
\label{eq:MHDfnc_dXi_x}
\eeq
where $b$ $(0<b<1)$ is a parameter for the transition width, 
and $c$ $(0 < c < 1)$ for the transition center.  
Also, its integral $\Xi(x;b,c)$ becomes 
\beq
\Xi(x;b,c) = \frac12\left[\,b\,\ln\cosh\left(\frac{x}{b}-c\right)+x \,\right]
+{\rm constant}. 
\label{eq:MHDfnc_Xi_x}
\eeq

In actual computations, functions $\Xi'(A_\phi)$ and $\Xi(A_\phi)$ are defined by
\beq
\Xi'(A_\phi)
= \frac12\left[\,\tanh\left(\frac1{b}
\frac{A_\phi-A_{\phi}^{0}}{A_\phi^{1} - A_{\phi}^{0}}
-c\right)+1\,\right],  
\label{eq:MHDfnc_dXi}
\eeq
and
\beqn
&&
\Xi(A_\phi)
\,=\, \frac12\left[\,b (A_\phi^{1}-A_{\phi}^{0})\times 
\phantom{\frac1{b}}\right.
\nonumber\\
&&\ \ \left.
\ln\cosh\left(
\frac1{b}
\frac{A_\phi-A_{\phi}^{0}}{A_\phi^{1} - A_{\phi}^{0}}
-c\right)+A_\phi\,\right]+{\rm constant},  \qquad
\label{eq:MHDfnc_Xi}
\eeqn
where the constant of Eq.~(\ref{eq:MHDfnc_Xi}) is set to be $\Xi(-\infty)=0$.
The function $\Xi'(A_\phi)$ varies from 0 to 1 in between 
$A_\phi^0 < A_\phi < A_\phi^{1}$.  We always set the value of 
$A_\phi$ at the rotation axis (z-axis) to be zero.

\subsubsection{Models}
\label{sec:models}

Using the function $\Xi'(A_\phi)$ and its integral $\Xi(A_\phi)$, 
we model the forms of arbitrary functions (\ref{eq:arbfnc}).  
For the EV-MT type solutions, we choose the same as Paper I, namely, 
\beqn
\Lambda &=& - \Lambda_0 \Xi(A_\phi) - \Lambda_1 A_\phi - {\cal E},
\label{eq:MHDfnc_Lambda}
\\[1mm]
A_t &=& - \Omegac A_\phi + C_e, 
\label{eq:MHDfnc_At}
\\[1mm]
&& \!\!\!\!\!\!\!\!\!\!\!\!\!\!\!\!
\sqrt{-g}\Lambda_\phi \,=\, \Lambda_{\phi 0}\, \Xi(A_\phi),
\label{eq:MHDfnc_Lambda_phi}
\\[1mm]
&& \!\!\!\!\!\!\!\!\!\!\!\!\!\!\!\!
\sqrt{-g}B \,=\, 4\pi \Lambda_{\phi 0}\, \Xi(A_\phi),
\label{eq:MHDfnc_B}
\eeqn
where $\Lambda_0$, $\Lambda_1$, ${\cal E}$, $\Omegac$, $C_e$, and 
$\Lambda_{\phi 0}$ are constant.  Values of 
$\Lambda_0$, $\Lambda_1$, and $\Lambda_{\phi 0}$ are prescribed 
to control the strength of electromagnetic fields, 
while those of ${\cal E}$, $\Omegac$, and $C_e$ are calculated 
from conditions to specify the mass, total angular momentum, and 
charge of a solution.

For computing solutions with the electromagnetic vacuum outside and with 
the force-free magnetotunnel (hereafter EV-MT type solutions), we choose, 
as in Paper I, the parameters $A_\phi^1$ and $A_\phi^0$ of the 
sigmoid functions $\Xi(A_\phi)$ appear in 
Eqs.~(\ref{eq:MHDfnc_Lambda}), (\ref{eq:MHDfnc_Lambda_phi}), and (\ref{eq:MHDfnc_B}) as 
$A_\phi^{1}=A_\phi^{\rm max}$ where $A_\phi^{\rm max}$ is 
the maximum value of $A_\phi$ on the stellar support, and 
$A_\phi^0 = A_{\phi, \rm S}^{\rm max}$ 
where $A_{\phi, \rm S}^{\rm max}$ is the maximum value of $A_\phi$ on 
the stellar surface.  
The choice $A_\phi^0 = A_{\phi, \rm S}^{\rm max}$ is necessary for 
the toroidal component of the magnetic fields to be confined inside of the star.

For computing solutions with the force-free magnetosphere 
the above choices for $A_\phi^0$ and $A_\phi^1$ are not necessary. 
In our calculation for the solutions with the force-free magnetosphere and 
the magnetotunnel (hereafter MS-MT type solutions), we choose $A_\phi^0$ and 
$A_\phi^1$ to be the same as above.  We also present the solutions 
whose $A_\phi^1$ is set larger than $A_\phi^{\rm max}$ and 
$A_{\phi}^{0}$ smaller than $A_{\phi, \rm S}^{\rm max}$ 
(see, Table \ref{tab:para_functions} below).  
In any case, note that $A_\phi^{\rm max} > A_{\phi, \rm S}^{\rm max} > 0$.

The relation (\ref{eq:MHDfnc_At}) implies that the star is uniformly 
rotating, $\Omega(A_\phi)=\Omegac$.  
Eq.~(\ref{eq:MHDfnc_B}) is an integral of Eq.~(\ref{eq:iMHD_FF}), 
where a constant of integration does not affect a solution.  
The value of $C_e$ is determined to set the net electric 
charge to vanish.

For the differentially rotating solutions, 
we modify the Eq.~(\ref{eq:MHDfnc_At})
\beq
A_t \,=\, - \Omegac \Xi(A_\phi) + C_e, 
\label{eq:MHDfnc_At_FF}
\eeq
and hence the rotation law becomes 
\beq
\Omega \,=\, \Omegac \Xi'(A_\phi).
\label{eq:MHDfnc_Omega_FF}
\eeq
This is a differential rotation law whose angular 
velocity $\Omega$ varies from 0 to $\Omegac$ 
as the $A_\phi$ from $A_\phi^0=0$ to $A_\phi^1$.  
We will explain this choice of differential rotation 
law and its parameters in the later section.

\subsection{Numerical computation}

\subsubsection{Setups for coordinate grids and multipoles}

The solutions presented below are associated 
with extremely strong mixed poloidal and toroidal magnetic fields.  
As in Paper I, our models produce mixed poloidal and toroidal 
fields concentrated near the equatorial surface.  Hence, it is 
necessary to resolve such configurations with a large number of 
grid points in $\theta$-coordinate, and accordingly, a large number 
of multipoles.  

The numbers of grid points and other grid parameters used 
in actual computations shown in the later sections are 
the same as the highest resolution used in Paper I, 
which is reproduced in Table \ref{tab:RNS_grids} marked as SE3tp.  
A system of equations is discretized on spherical coordinates 
$(r,\theta,\phi)\in [r_a,r_b]\times[0,\pi]\times[0,2\pi]$ 
where $r_a=0$ is the center of the star, and $r_b=10^6 R_0$, 
where $R_0$ is the equatorial radius of the star.  
To resolve a toroidal region of extremely strong magnetic 
fields concentrated near the equatorial surface, we include 
the multipoles up to $L=60$.  Details of convergence tests 
can be also found in Paper I.  

\begin{table}
\caption{Grid parameters used for computing magnetized 
rotating compact stars.  
Normalized radial coordinates $r_a$, $r_b$, and $r_c$ are in the 
unit of equatorial radius $R_0$ in the coordinate length.
}
\label{tab:RNS_grids}
\begin{tabular}{cccccccccl}
\hline
\qquad Type  & $r_a$ & $r_b$ & $r_c$ & $\Nrf$ & $\Nrm$ & $N_r$ & $N_\theta$ & $N_\phi$ & $L$ \\
\qquad SE3tp & 0.0 & $10^6$ & 1.1  & 160 & 176  & 384  & 384  & 72 & 60 \\
\hline
\multicolumn{10}{l}{\,$r_a$ : Radial coordinate where the radial grids starts.} \\
\multicolumn{10}{l}{\,$r_b$ : Radial coordinate where the radial grids ends.}     \\
\multicolumn{10}{l}{\,$r_c$ : Radial coordinate between $r_a$ and $r_b$ where}   \\
\multicolumn{10}{l}{\qquad the radial grid spacing changes.}   \\
\multicolumn{10}{l}{$N_r$\ \ : Number of intervals $\Dl r_i$ in $r \in[r_a,r_b]$.} \\
\multicolumn{10}{l}{$\Nrf$\ \ : Number of intervals $\Dl r_i$ in $r \in[r_a,1]$.} \\
\multicolumn{10}{l}{$\Nrm$ : Number of intervals $\Dl r_i$ in $r \in[r_a,r_{c}]$.} \\
\multicolumn{10}{l}{$N_{\theta}$\ \ : Number of intervals $\Dl \theta_j$ in $\theta\in[0,\pi]$.} \\
\multicolumn{10}{l}{$N_{\phi}$\ \ : Number of intervals $\Dl \phi_k$ in $\phi\in[0,2\pi]$.} \\
\multicolumn{10}{l}{$L$\ \ \ \ : Order of included multipoles.} \\
\hline
\end{tabular}  
\end{table}

\subsubsection{Model parameters}

Configuration and intensity of electromagnetic fields inside and outside of 
the compact stars are determined by the forms of arbitrary functions presented 
in Sec.~\ref{sec:models}, and parameters associated with them.  
The parameters used in the present computations, that is, 
the parameters $b$ and $c$ defined in Eqs.~(\ref{eq:MHDfnc_Lambda}), 
(\ref{eq:MHDfnc_Lambda_phi}), and (\ref{eq:MHDfnc_B}),  
and those for differential rotation in (\ref{eq:MHDfnc_At_FF}) and 
(\ref{eq:MHDfnc_Omega_FF}) are all listed in Table \ref{tab:para_functions}.  
These functions and parameters 
produce extremely strong mixed poloidal and toroidal magnetic fields.  
In particular, the values of parameters are close but changed 
from those of Paper I that the electromagnetic fields become 
stronger enough to form a toroidal force-free magnetotunnel.

For the equations of states (EOS), we choose a 
polytropic EOS 
\beq
p\,=\,K \rho^\Gamma, 
\label{eq:EOS}
\eeq
for simplicity.  This introduce two more parameters, 
a polytropic constant $K$ and index $\Gamma$, 
whose values are set as in Table \ref{tab:TOV_solutions}. 

In addition, we have three parameters $\{{\cal E}, \Omegac, C_e \}$ 
in Eqs.~(\ref{eq:MHDfnc_Lambda}) and (\ref{eq:MHDfnc_At}) 
(or (\ref{eq:MHDfnc_At_FF}) for differentially rotating model) and 
one augmented parameter $R_0$.  The parameter $C_e$ does not change 
the solution when the star is surrounded by the magnetosphere, and so 
$C_e$ may be set to zero in this case.  For the case with electromagnetic 
vacuum outside, $C_e$ is set for an asymptotic electric charge 
\beq
Q\,=\, \frac1{4\pi}\int_\infty \Fabu dS_{\albe}, 
\label{eq:charge}
\eeq
to be zero.  
The other three parameters calculated from three conditions, 
setting the value of the rest mass density at the center $\rho_c$, 
the normalization of the equatorial radius, $r_{\rm eq}$, as 
$r_{\rm eq}/R_0=1$, and the value of the radius along 
the rotation axis $r_p/r_{\rm eq}$.  
Further details can be found in Paper I.  

\begin{table}
\caption{Parameters related to arbitrary functions in the integrability 
conditions (\ref{eq:MHDfnc_Lambda})--(\ref{eq:MHDfnc_Omega_FF}), 
used in computing solutions presented in Sec.~\ref{sec:Results}.  
The parameters $(b,c)$ are those used in 
the sigmoid function $\Xi'(A_\phi)$ for Eqs.~(\ref{eq:MHDfnc_Lambda}), 
(\ref{eq:MHDfnc_Lambda_phi}) and (\ref{eq:MHDfnc_B}), while DR:$(b, c)$ 
are used for the differential rotation in Eqs.~(\ref{eq:MHDfnc_At_FF}) and 
(\ref{eq:MHDfnc_Omega_FF}).
The values of $A_\phi^0$ and $A_\phi^1$ in Eqs.~(\ref{eq:MHDfnc_Lambda}), 
(\ref{eq:MHDfnc_Lambda_phi}) and (\ref{eq:MHDfnc_B}) are set as below, and 
for DR models in Eqs.~(\ref{eq:MHDfnc_At_FF}) and 
(\ref{eq:MHDfnc_Omega_FF}), we set $A_\phi^1$ to be the same as the other 
$\Xi'(A_\phi)$, but set $A_\phi^0=0$.
}
\label{tab:para_functions}
\begin{tabular}{cccccccc}
\hline
Models & $\ \ \Lambda_0$ & $\Lambda_1$ & $\Lambda_{\phi0}$ & $(b, c)$ & DR:$(b, c)$ 
& $\dfrac{A_\phi^0}{A_{\phi, \rm S}^{\rm max}}$ & $\dfrac{A_\phi^1}{A_\phi^{\rm max}}$\\
\hline
EV-MT-UR & $-4.8$ & $0.3$ & $3.2$ & $(0.2, 0.5)$ &  ---  & $1.0$ & $1.0$ \\
MS-MT-DR & $-6.0$ & $0.3$ & $3.8$ & $(0.2, 0.5)$ & $(0.05, 0.25)$ & $1.0$ & $1.0$ \\
EV-MT-DR & $-6.0$ & $0.3$ & $3.8$ & $(0.2, 0.5)$ & $(0.1,  0.3)$  & $1.0$ & $1.0$ \\
MS-DR-1  &$\ \ 4.8$&$0.3$ & $1.2$ & $(0.2, 0.5)$ & $(0.02, 0.1)$  & $0.3$ & $1.7$ \\
MS-DR-2  & $-1.3$ & $0.3$ & $1.1$ & $(0.2, 0.5)$ & $(0.02, 0.1)$  & $0.3$ & $1.7$ \\
\hline
\end{tabular}
\end{table}

\begin{table}
\caption{
Quantities at the maximum mass model of Tolman-Oppenheimer-Volkoff (TOV) 
solutions with the polytropic EOS (\ref{eq:EOS}) in $G=c=M_\odot=1$ units.  
$p_c$ and $\rho_c$ are the pressure and the rest mass density at the center, 
$M_0$ is the rest mass, $M$ the gravitational mass, and 
$M/R$ the compactness (a ratio of the gravitational mass to 
the circumferential radius).  
The polytropic constant $K$ is chosen so that the value of $M_0$ 
becomes $M_0=1.5$ at the compactness $M/R=0.2$.  
To convert a unit of $\rhoc$ to cgs, multiply the values by 
$\Msol(G\Msol/c^2)^{-3} \approx 6.176393\times 10^{17} {\rm g}\ {\rm cm}^{-3}$.}  
\label{tab:TOV_solutions}
\begin{tabular}{cccccc}
\hline
$\Gamma$ & $\prhoc$ & $\rhoc$ & $M_0$ & $M$ & $M/R$ \\
\hline
$2$ & $0.318244$ & $0.00448412$ & $1.51524$ & $1.37931$ & $0.214440$\\
\hline
\end{tabular}
\end{table}
\begin{figure*}
\begin{center}
\includegraphics[height=42mm]{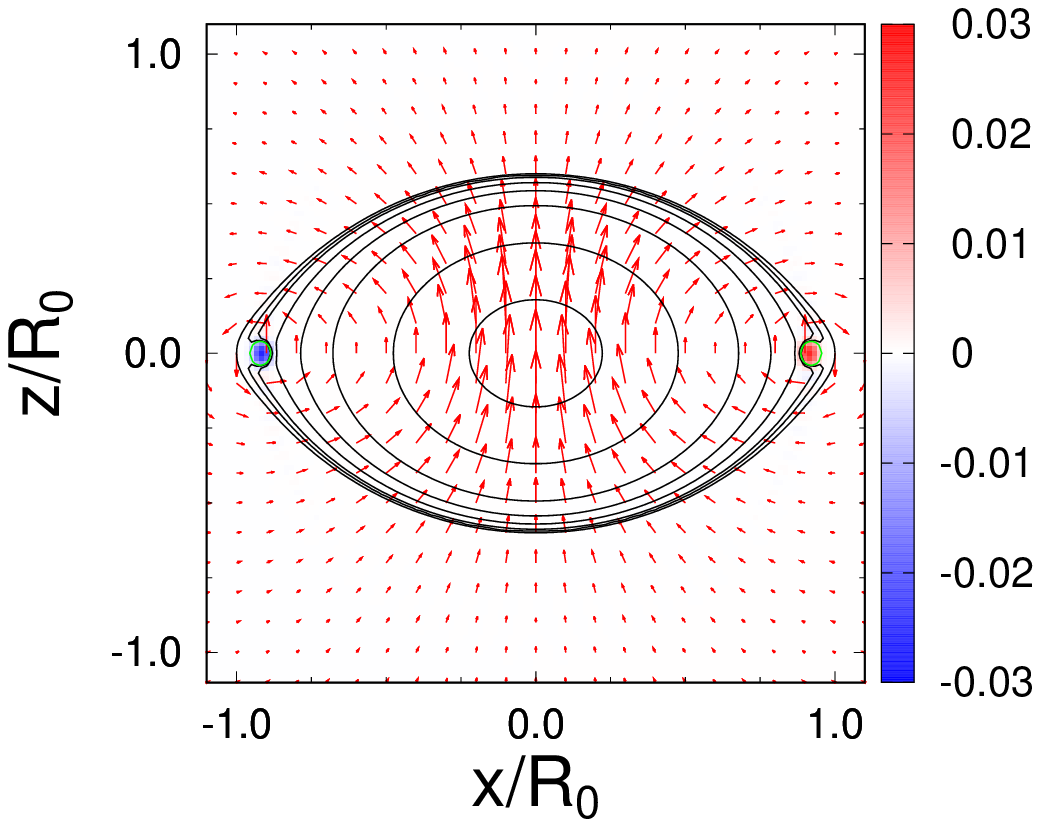}
\includegraphics[height=42mm]{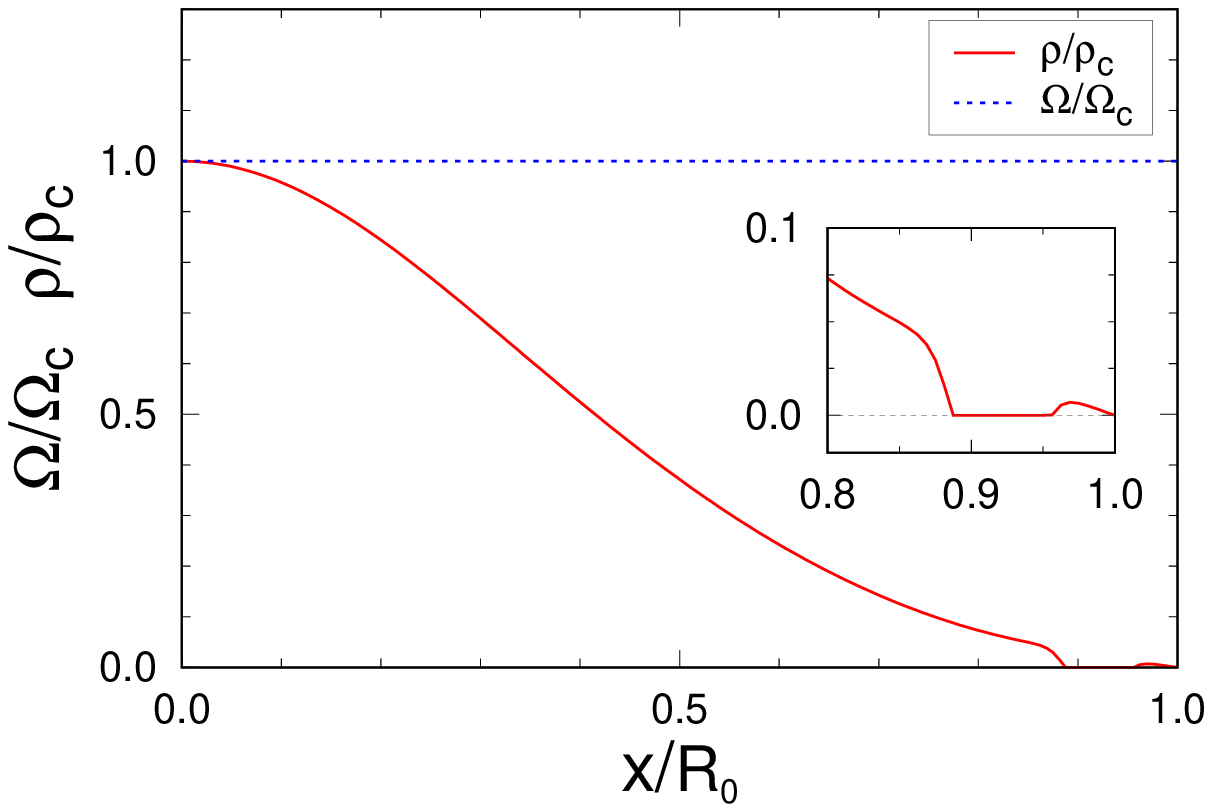}
\includegraphics[height=42mm]{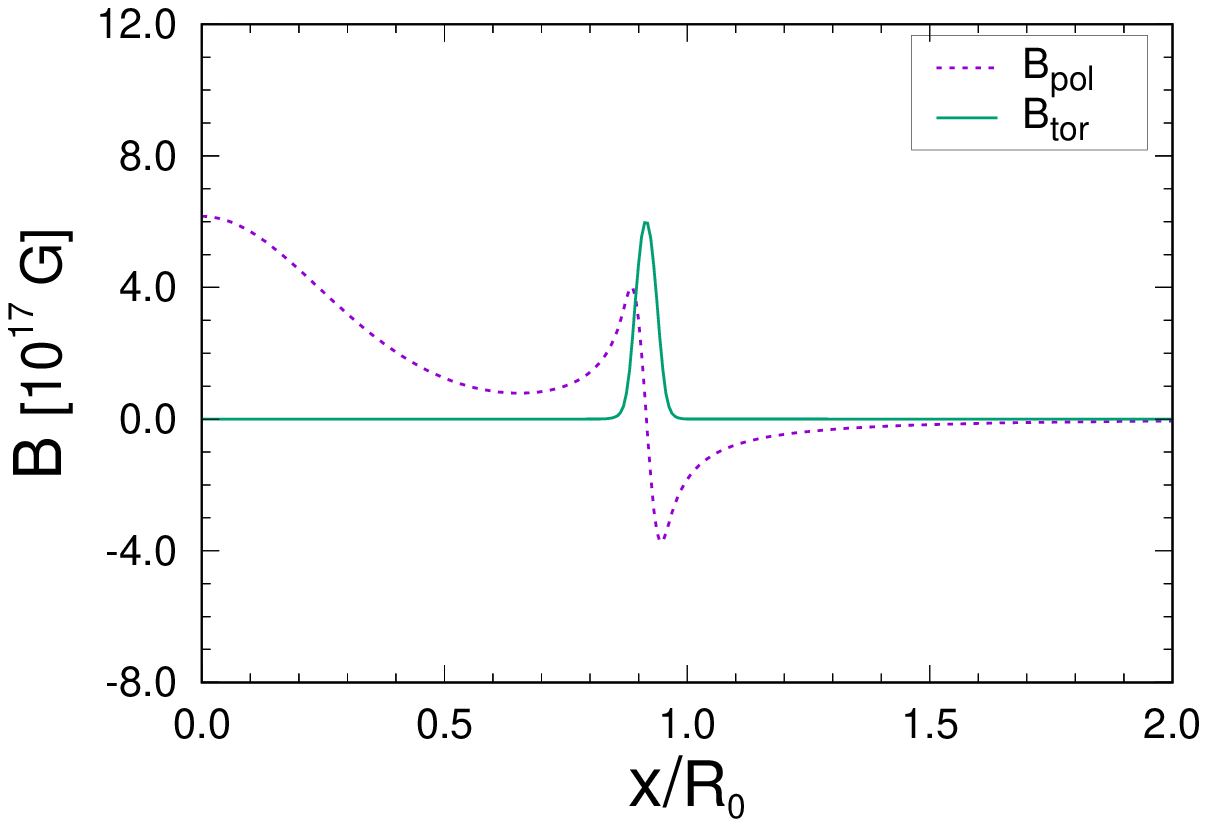}
\\
\includegraphics[height=42mm]{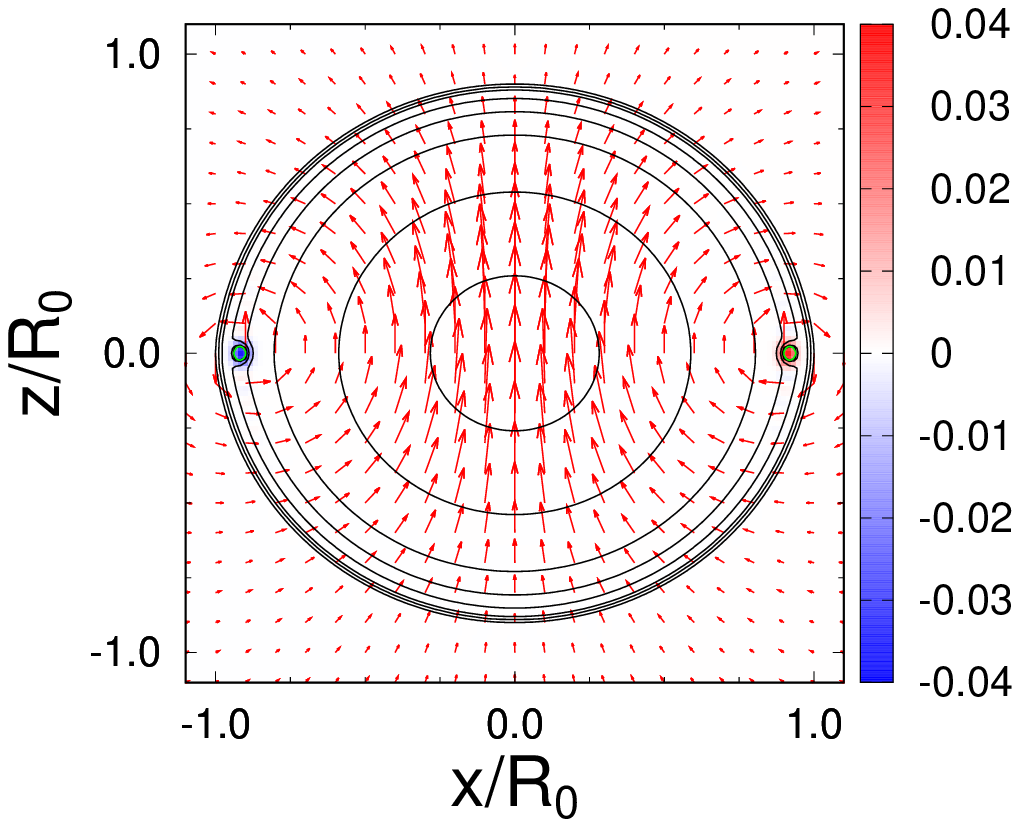}
\includegraphics[height=42mm]{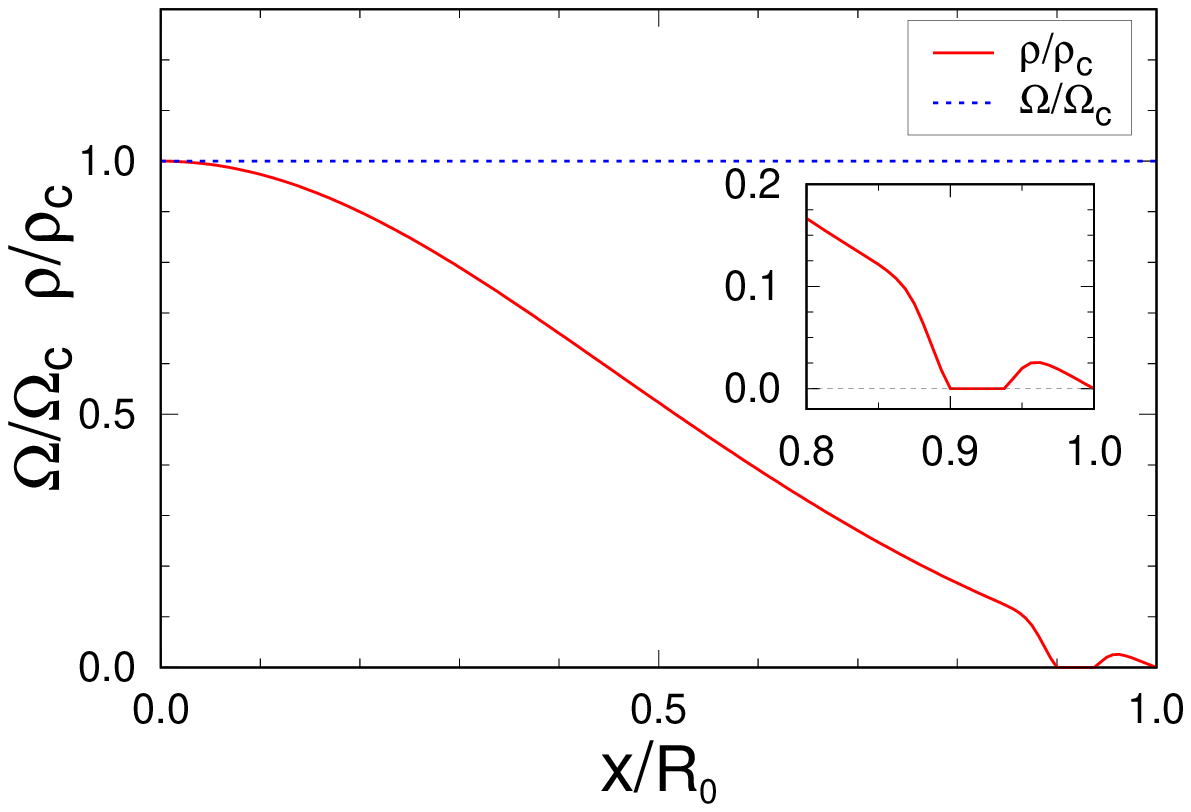}
\includegraphics[height=42mm]{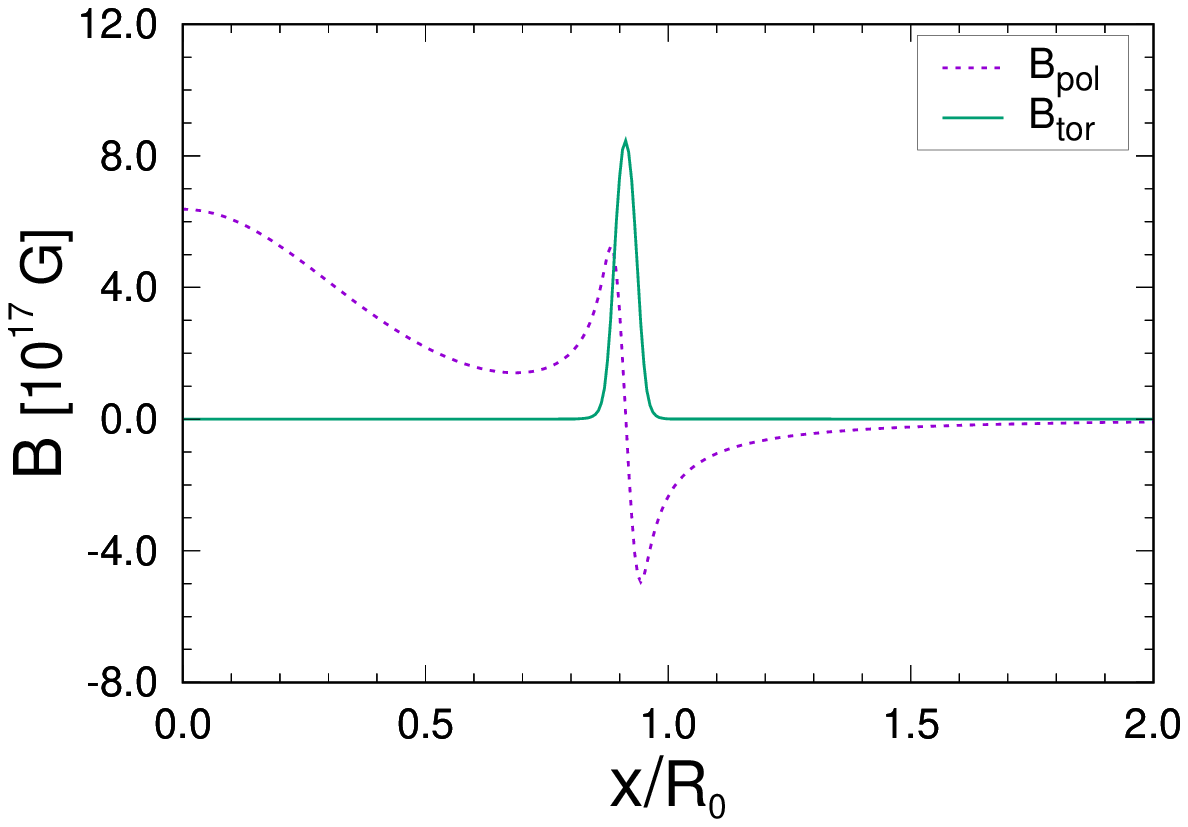}
\\
\includegraphics[height=36mm]{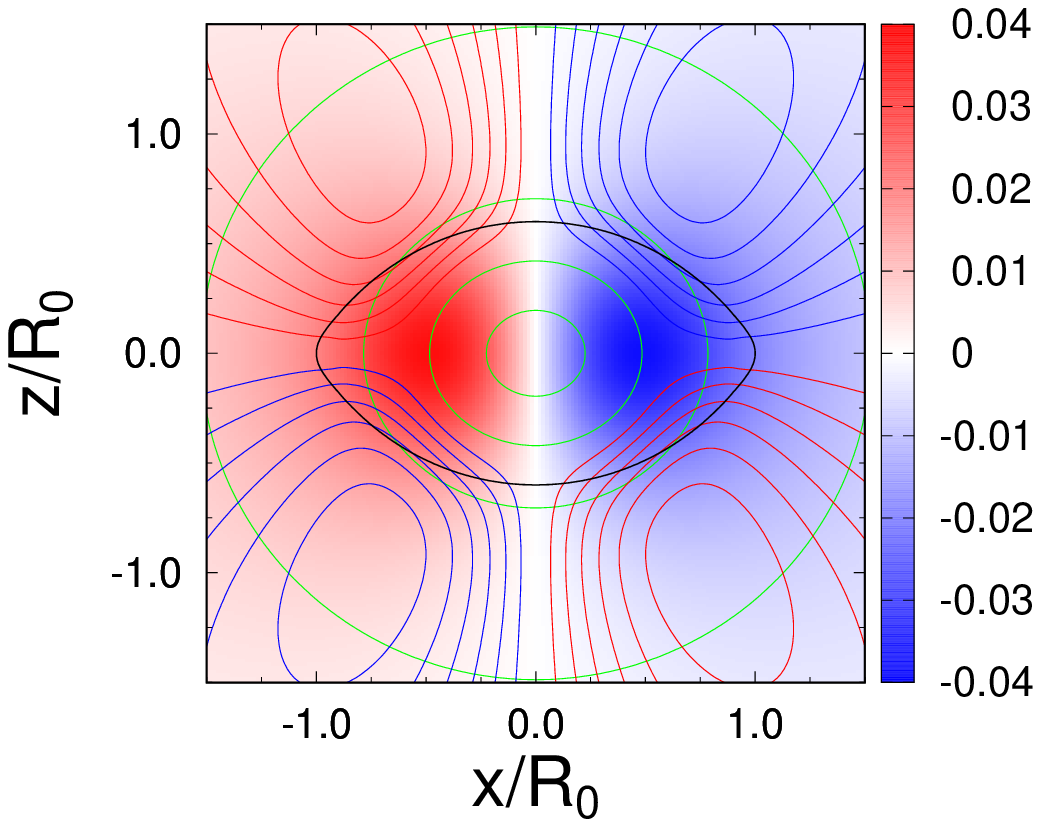}
\includegraphics[height=36mm]{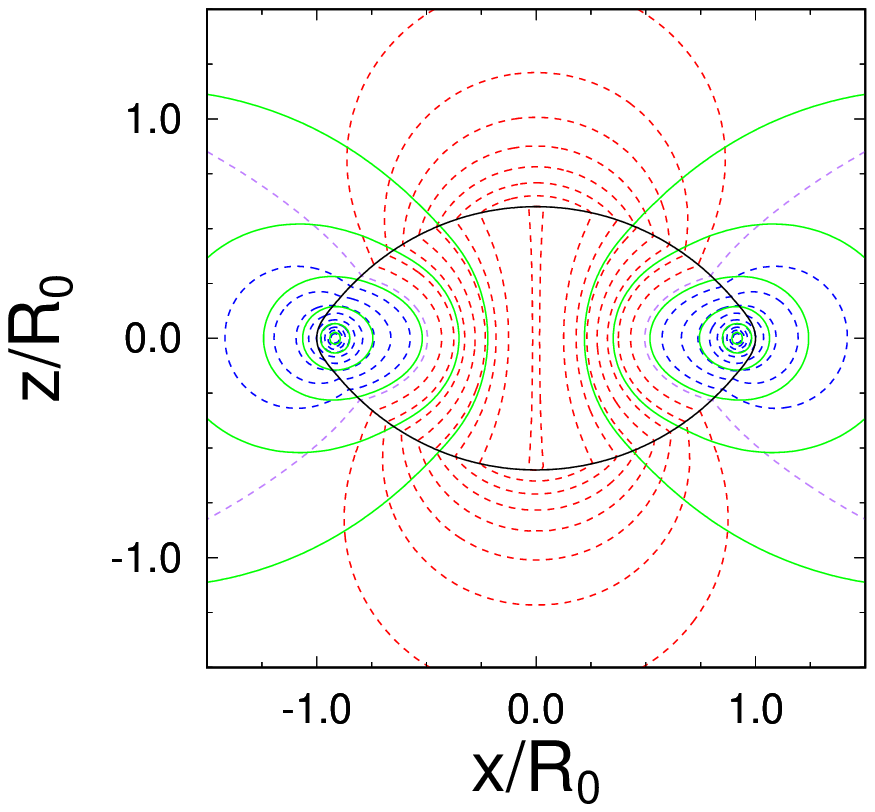}
\includegraphics[height=36mm]{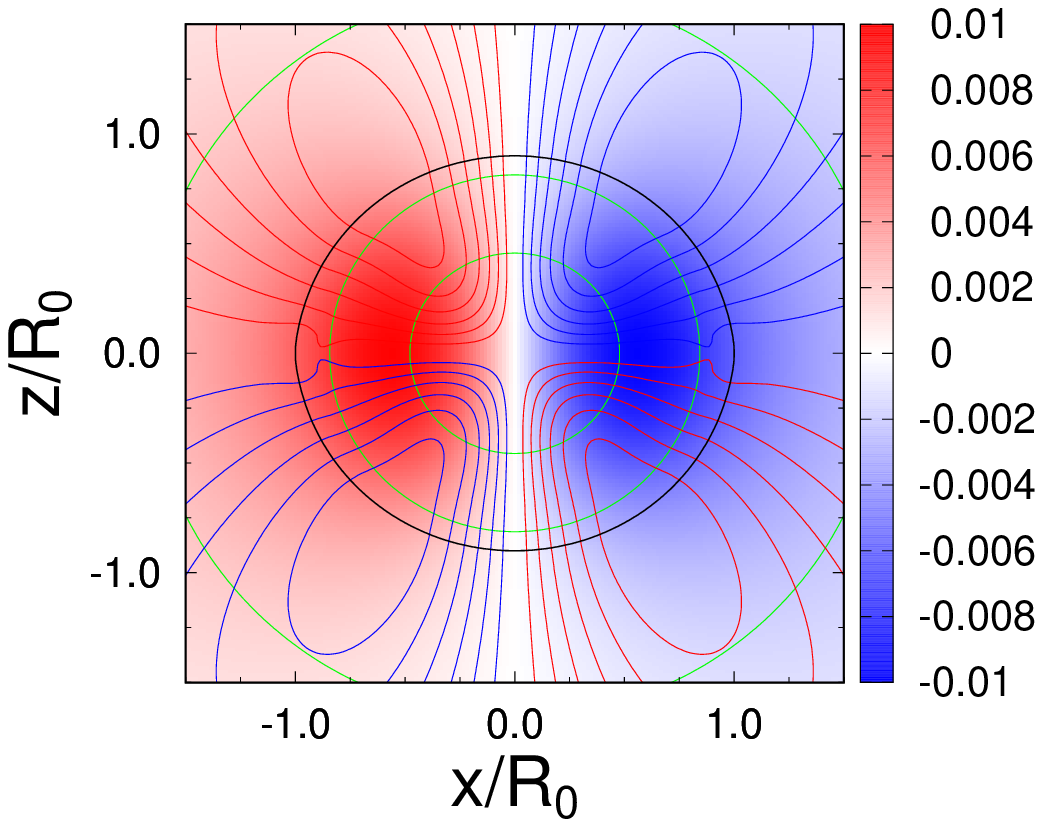}
\includegraphics[height=36mm]{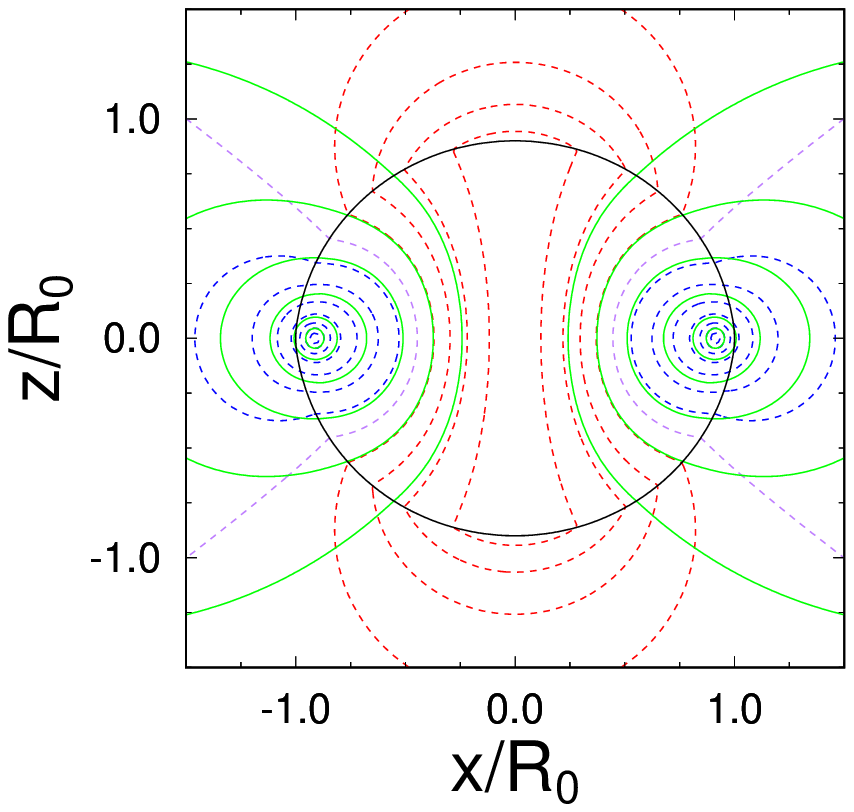}
\\
\caption{
Solutions for uniformly rotating extremely magnetized compact stars associated 
with an electromagnetic vacuum outside and magnetotunnel (EV-MT-UR type).  
The panels in the first row correspond to the rapidly rotating model EV-MT-UR-1.  
The left panel: contours of $p/\rho$ (black closed curves), 
the poloidal magnetic field (orange arrows), color density map 
for the toroidal magnetic fields (red and blue), and the boundary of 
the magnetotunnel (green circles) are shown.  The contours of $p/\rho$ 
are drawn at $p/\rho=0.001,0.002,0.005,0.01,0.02,0.05,0.1$.  
The middle panel: the rest mass density $\rho/\rho_c$ (red curve) and 
the angular velocity $\Omega/\Omega_c$ are plotted along the 
equatorial radius ($x$-axis).  
An inset is a close-up of $\rho/\rho_c$ near the surface.  
The right panel: components of the magnetic fields, $B_{\rm pol} = F_{xy}$ 
(dashed purple curve) and $B_{\rm tor} = -F_{xz}$ (dark green curve) 
are plotted along the equatorial radius ($x$-axis).  
The panels in the second row are the same as the first row but for 
the slowly rotating model EV-MT-UR-2.  
In the third row, the first panel from the left, the metric 
potentials are shown, which are the contours of $\psi$ (green closed curves), 
the color density map for $\tbeta_y$ (red and blue), 
the contours of $h_{xz}$ (red and blue curves), and the surface of 
the star (black closed curve).  In the second panel from the left, 
the components of electromagnetic 1-form are shown, which are the contours of 
$A_\phi$ (green curves), the contours of $A_t$ (dashed red (positive), 
purple (zero), blue (negative)), and the surface of the star 
(black closed curve) for the model EV-MT-UR-1.  
The third and fourth panels of the third row are the same as the 
first and the second panels, respectively, but for the model EV-MT-UR-2.  
}  
\label{fig:EV-MT-UR}
\end{center}
\end{figure*}

\begin{figure*}
\begin{center}
\includegraphics[height=42mm]{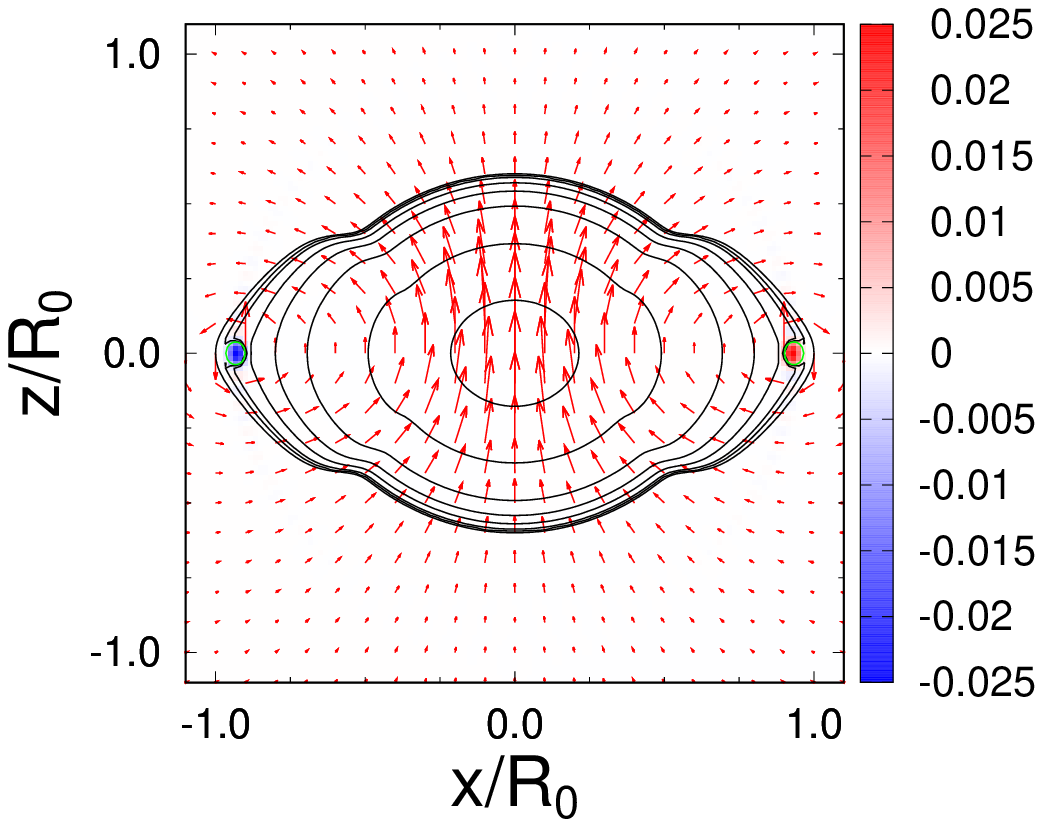}
\includegraphics[height=42mm]{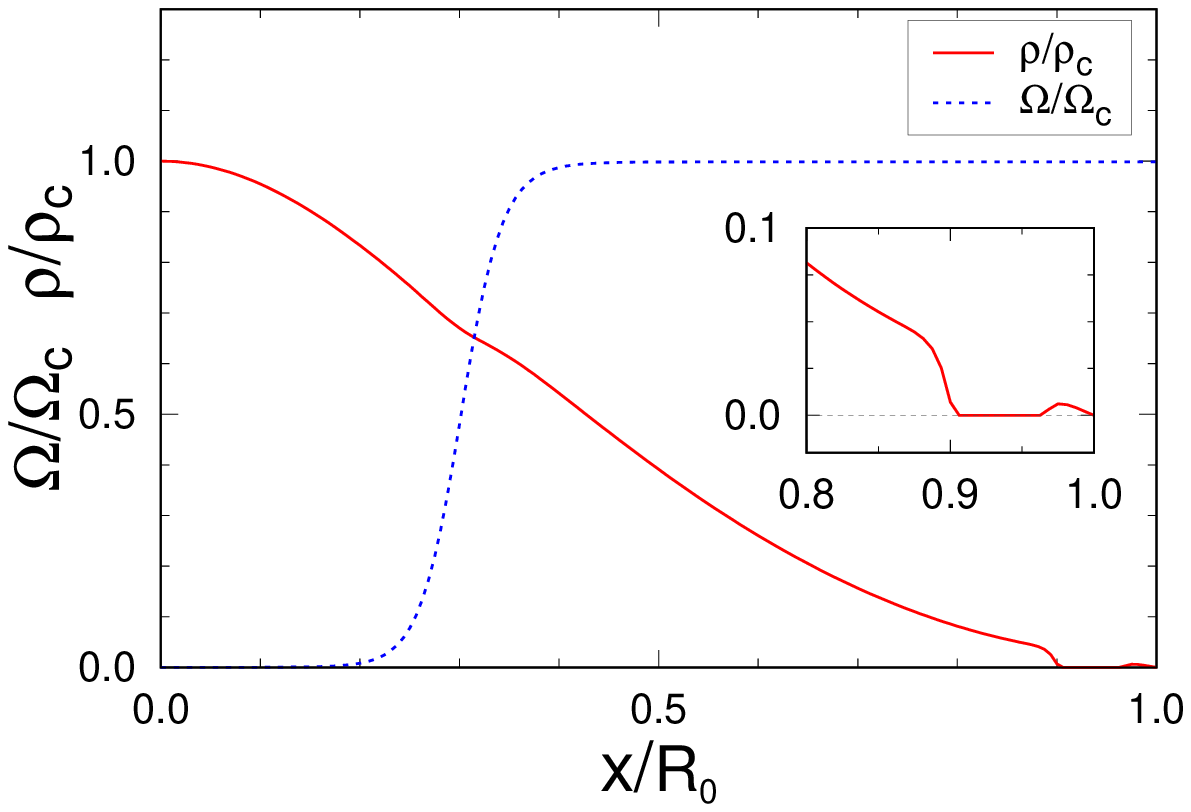}
\includegraphics[height=42mm]{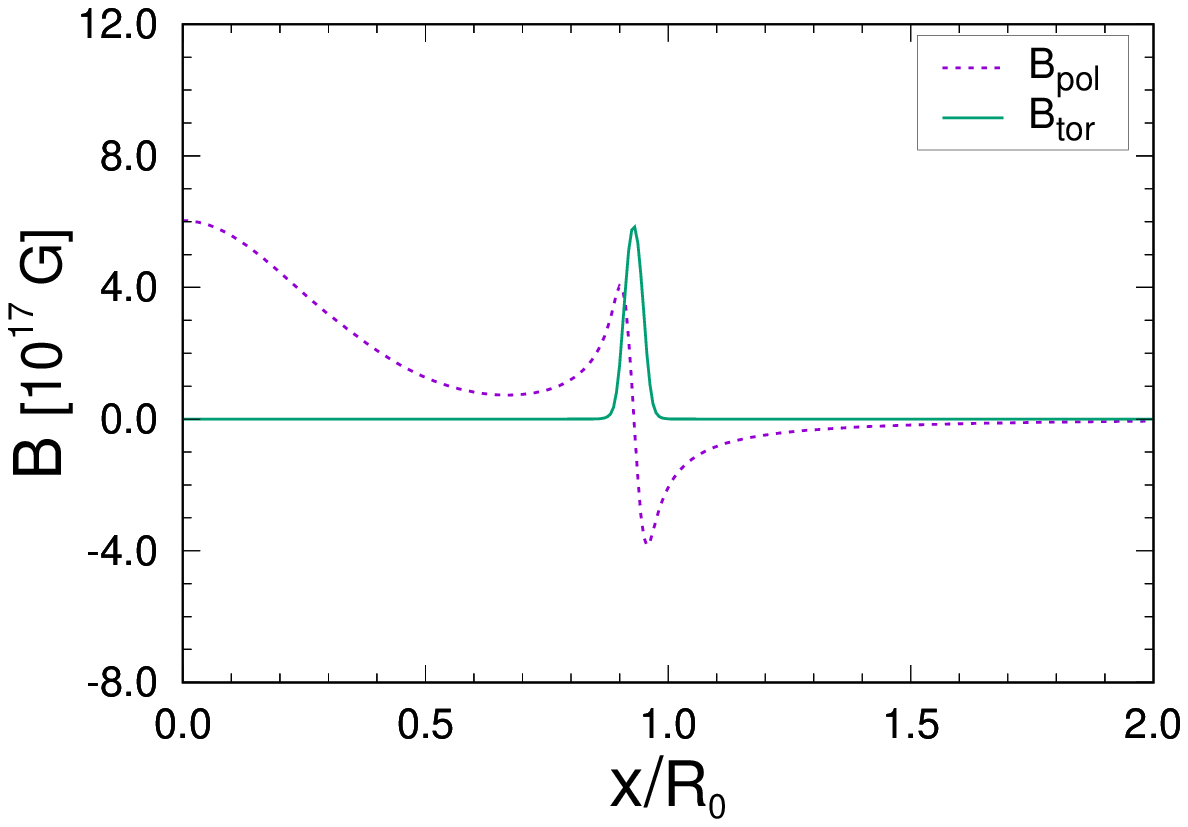}
\\
\includegraphics[height=42mm]{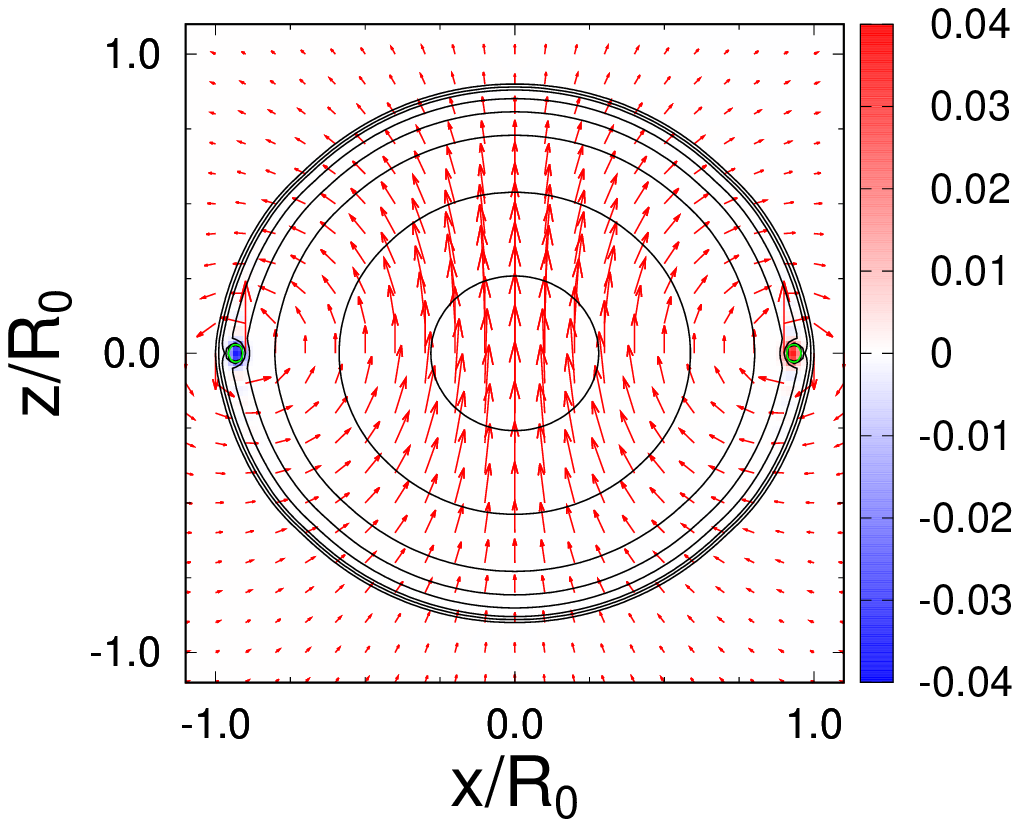}
\includegraphics[height=42mm]{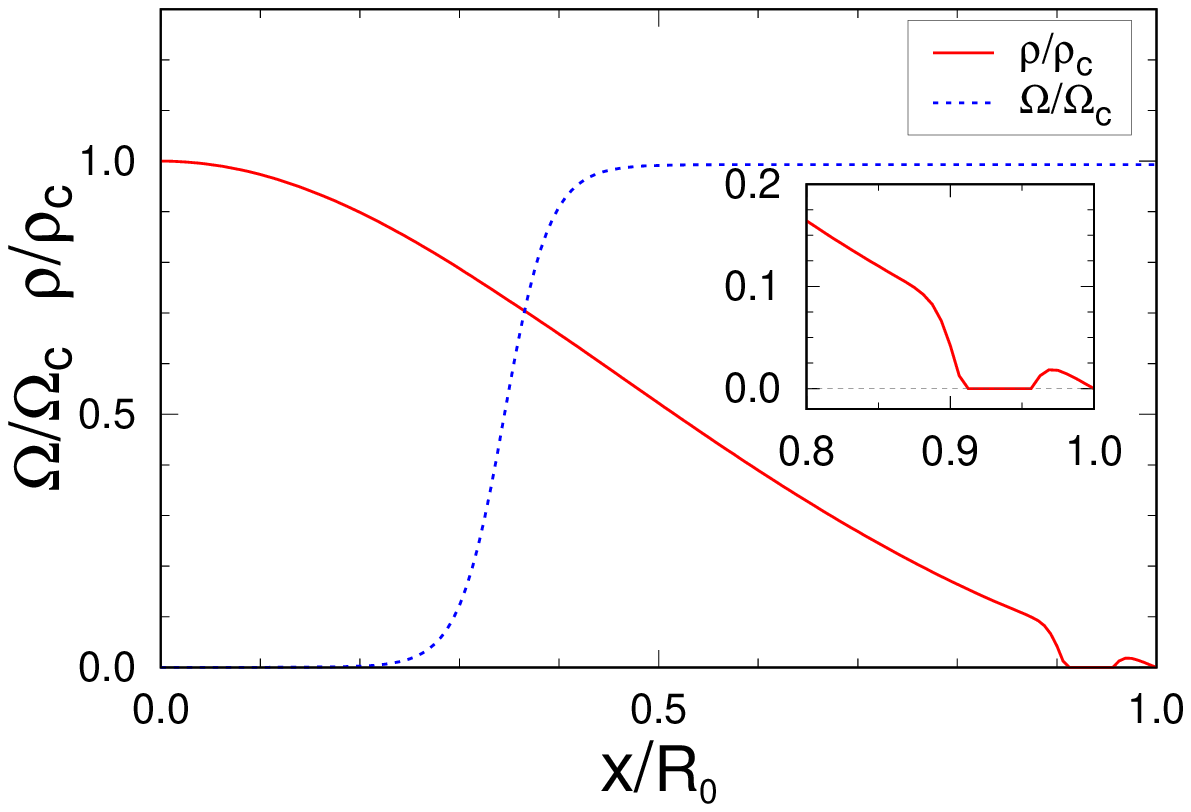}
\includegraphics[height=42mm]{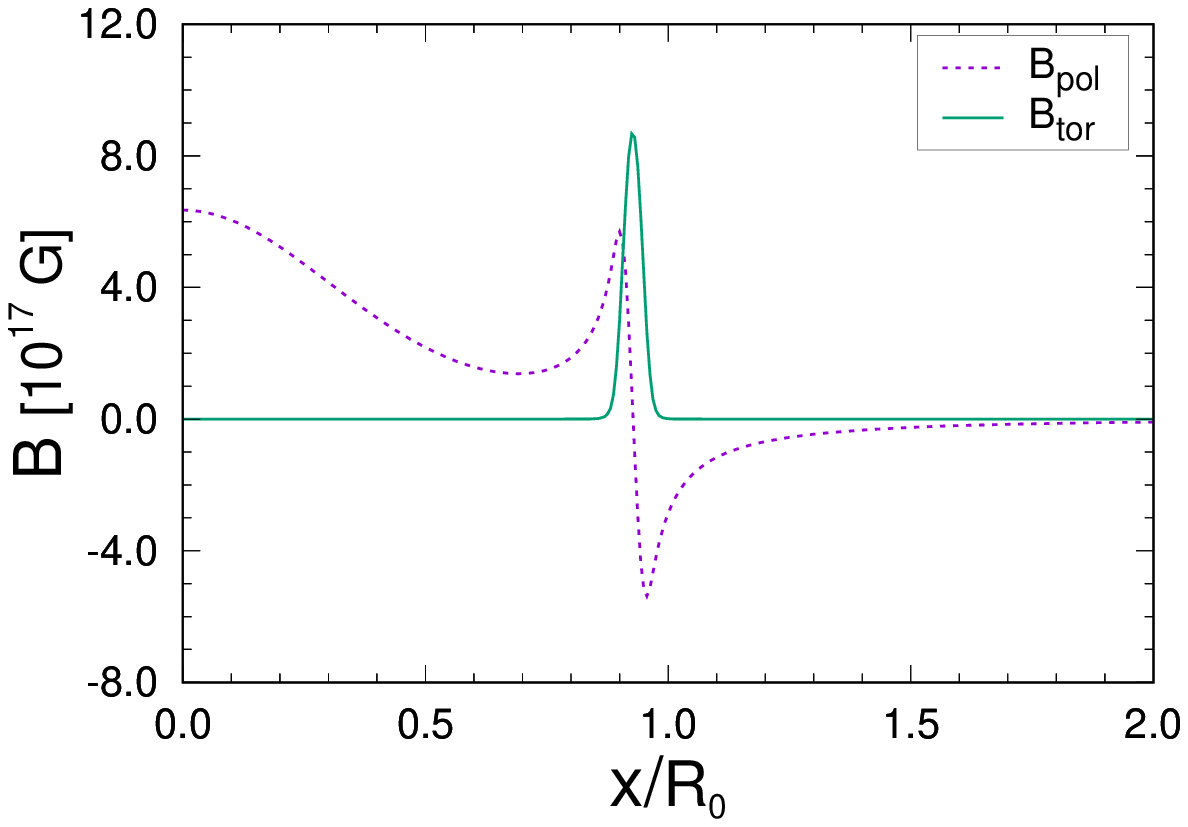}
\\
\includegraphics[height=36mm]{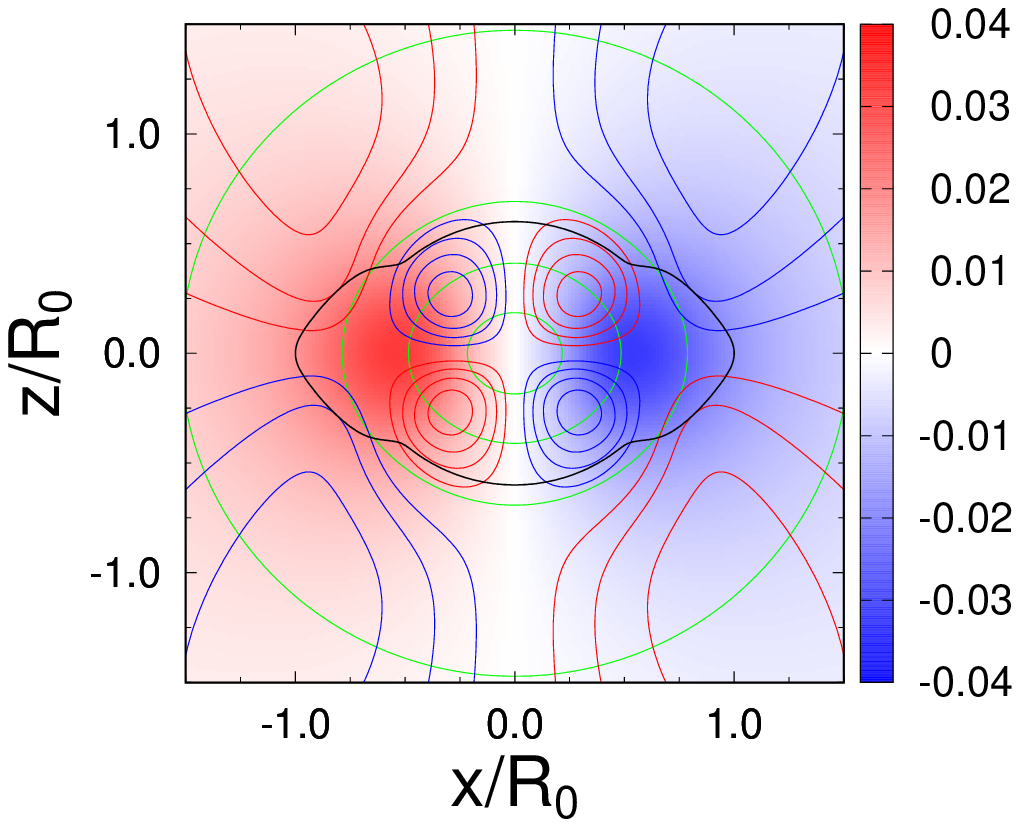}
\includegraphics[height=36mm]{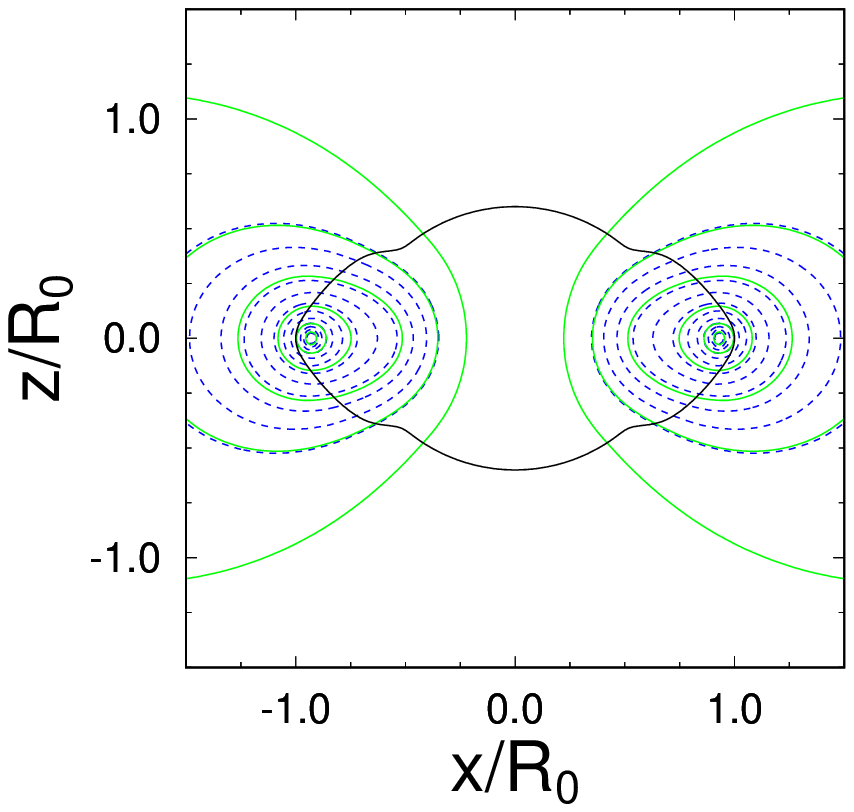}
\includegraphics[height=36mm]{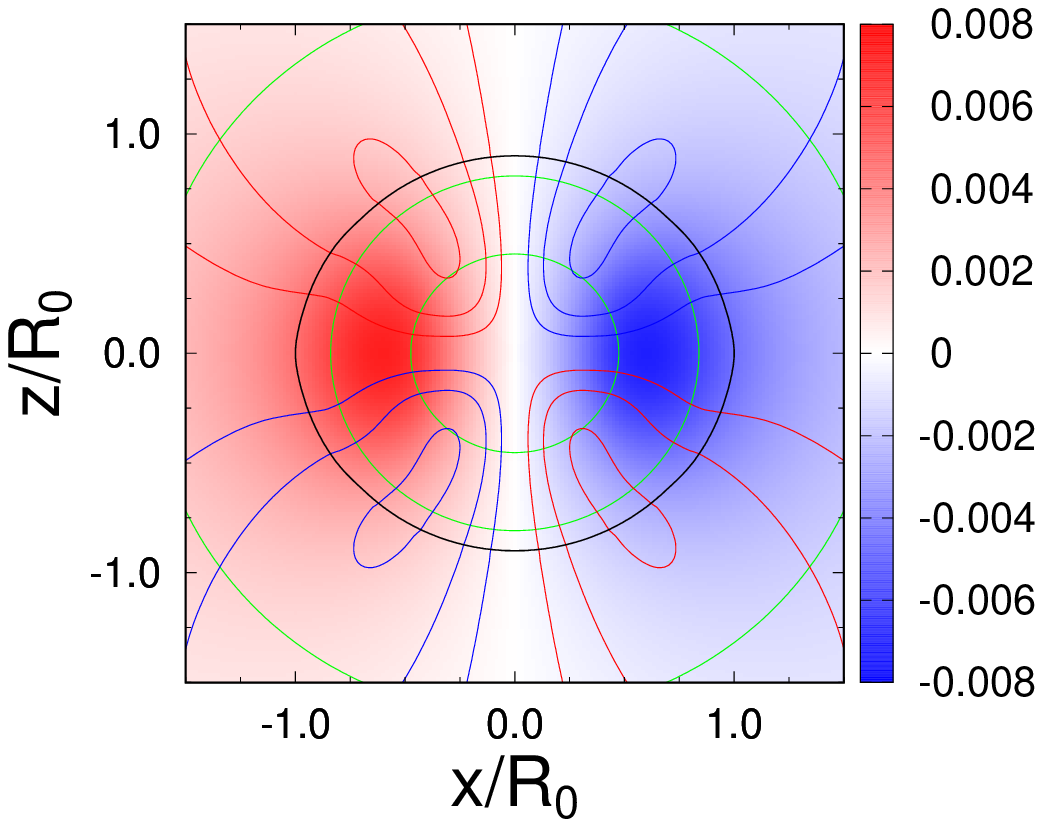}
\includegraphics[height=36mm]{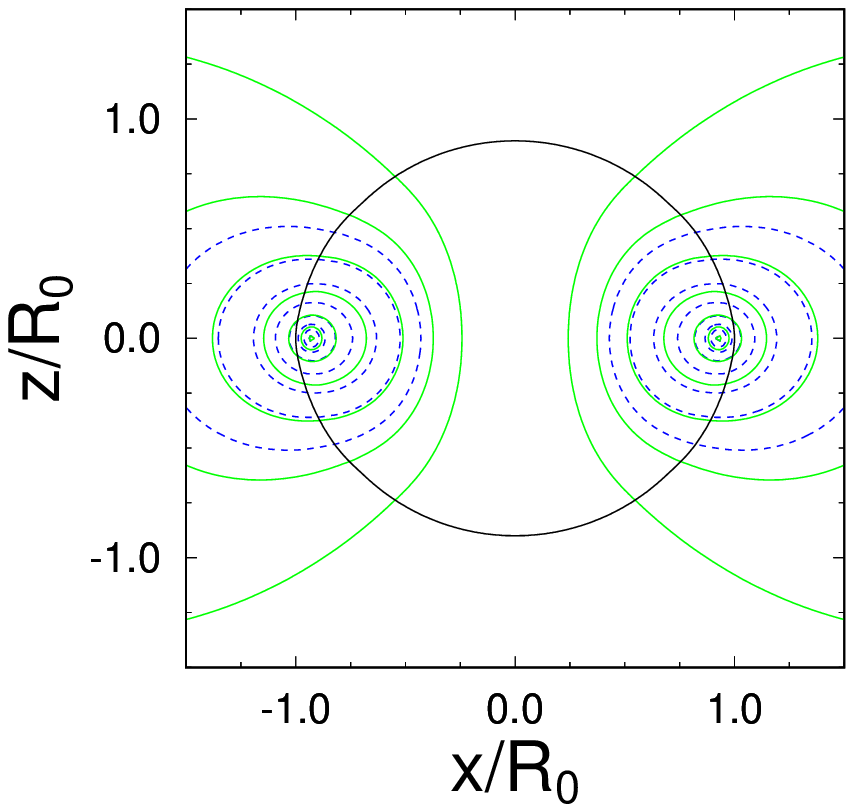}
\caption{
Same as Fig.\ref{fig:EV-MT-UR} but for differentially rotating 
and extremely magnetized compact stars 
associated with a magnetosphere and a magnetotunnel, MS-MT-DR-1 
(rapidly rotating model) and MS-MT-DR-2 (slowly rotating model).  
}  
\label{fig:MS-MT-DR}
\end{center}
\end{figure*}

\begin{figure*}
\begin{center}
\includegraphics[height=42mm]{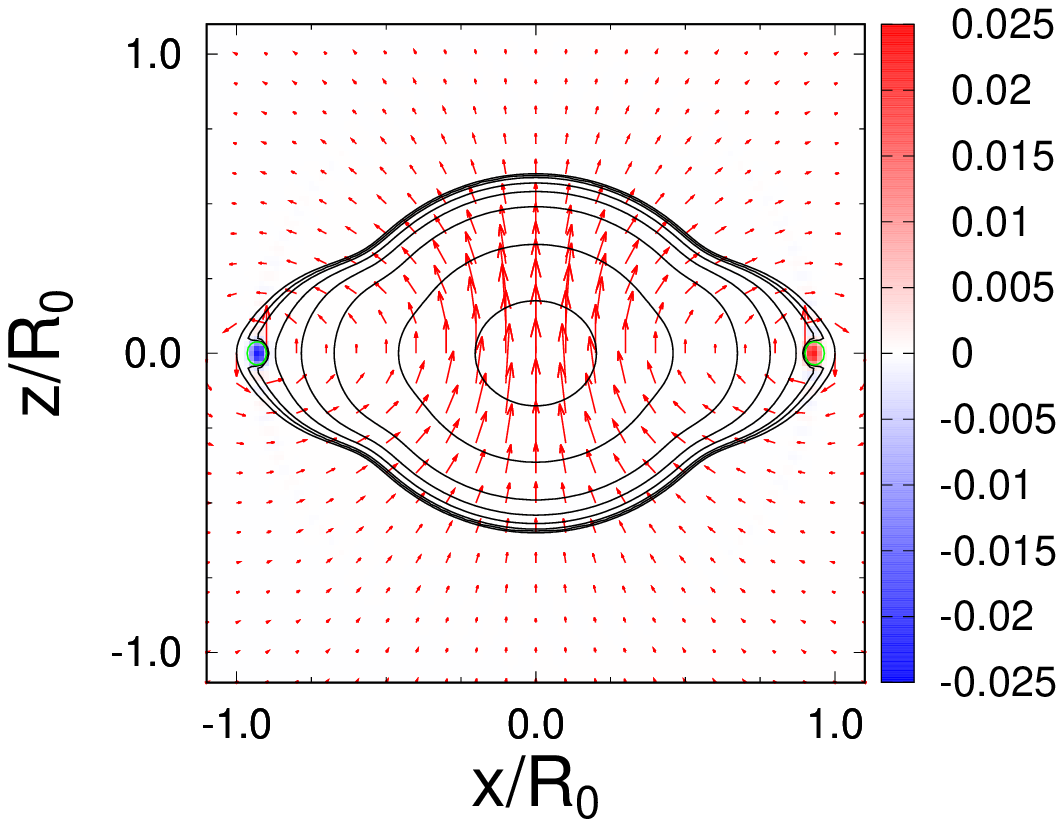}
\includegraphics[height=42mm]{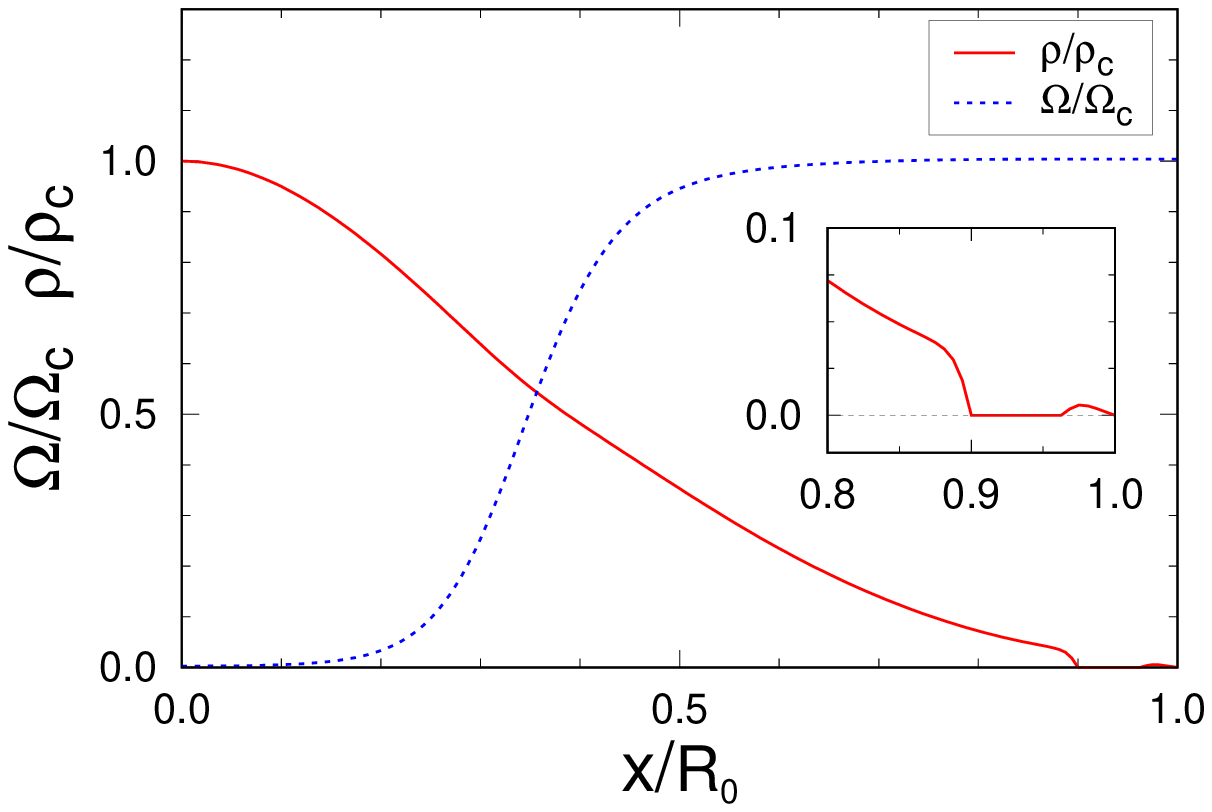}
\includegraphics[height=42mm]{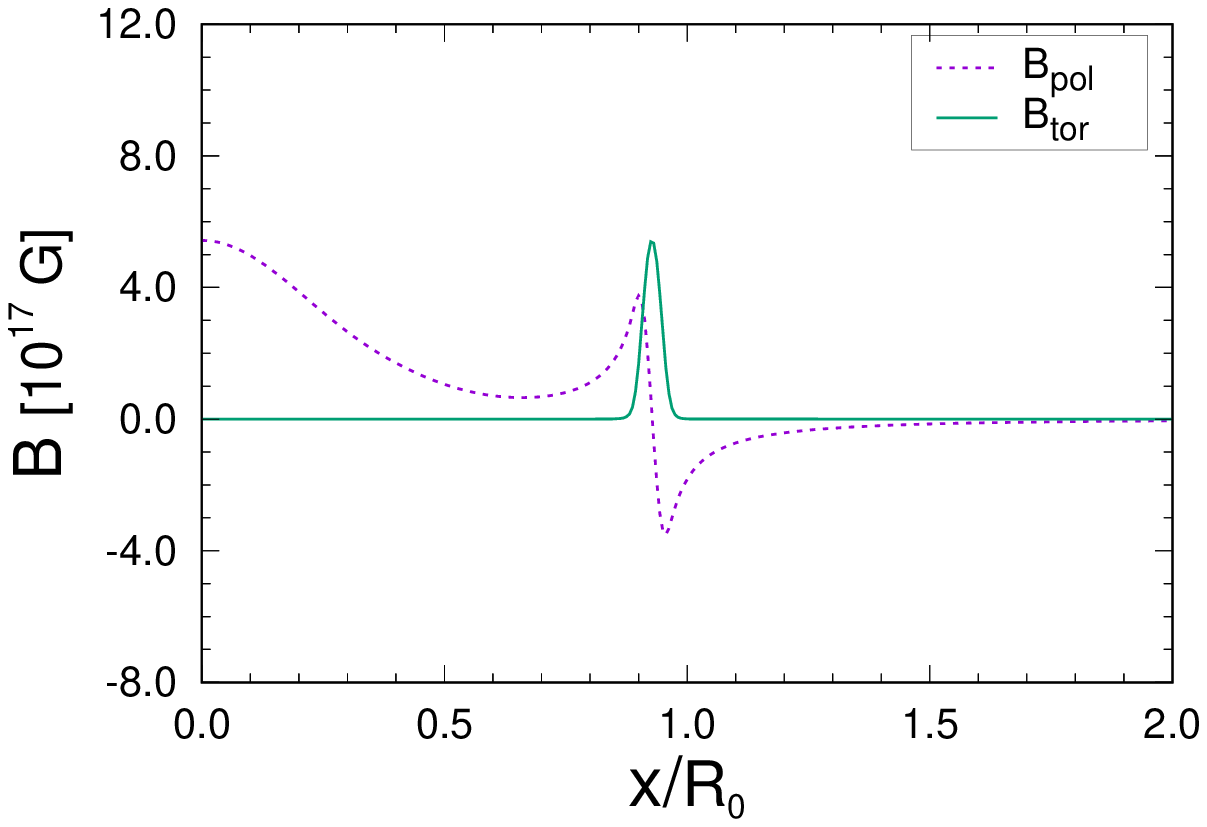}
\\
\includegraphics[height=42mm]{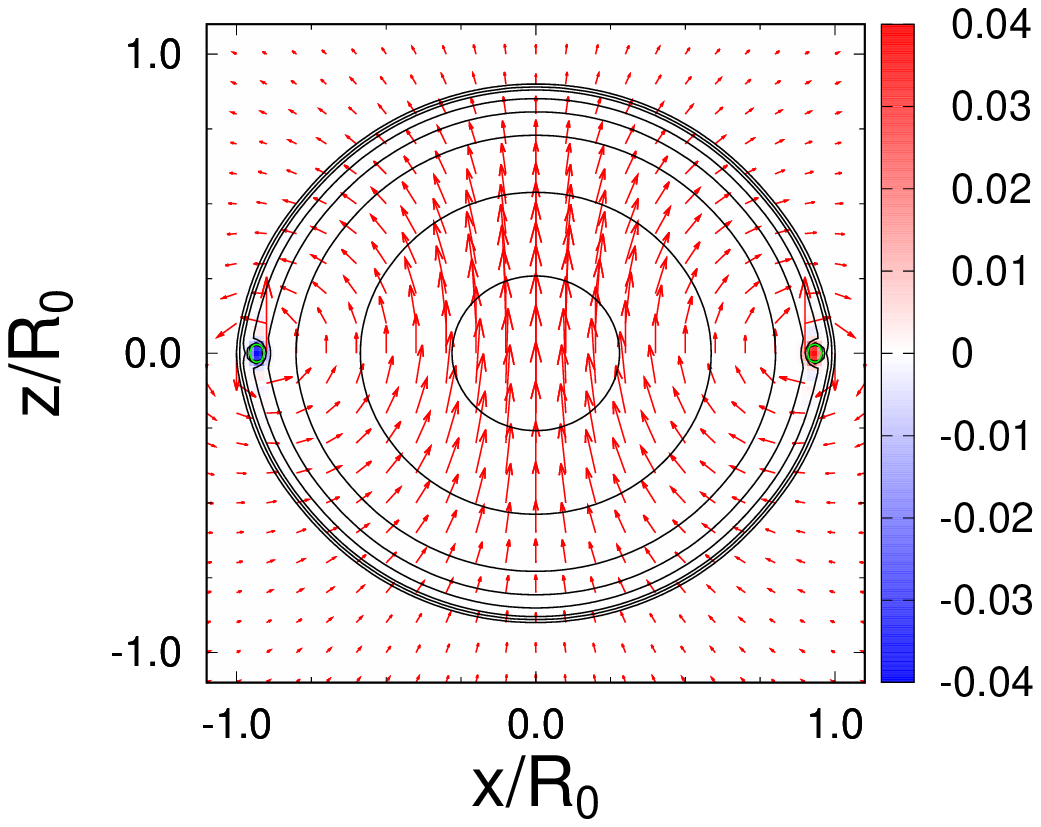}
\includegraphics[height=42mm]{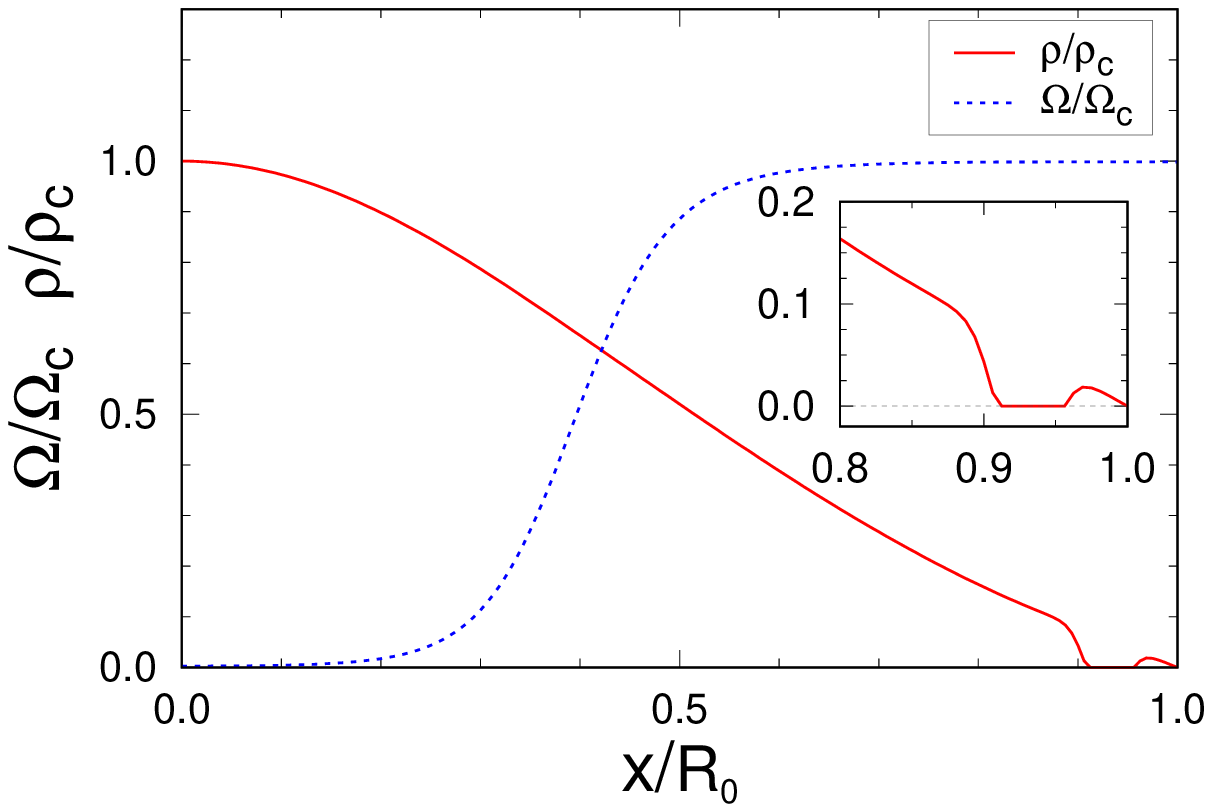}
\includegraphics[height=42mm]{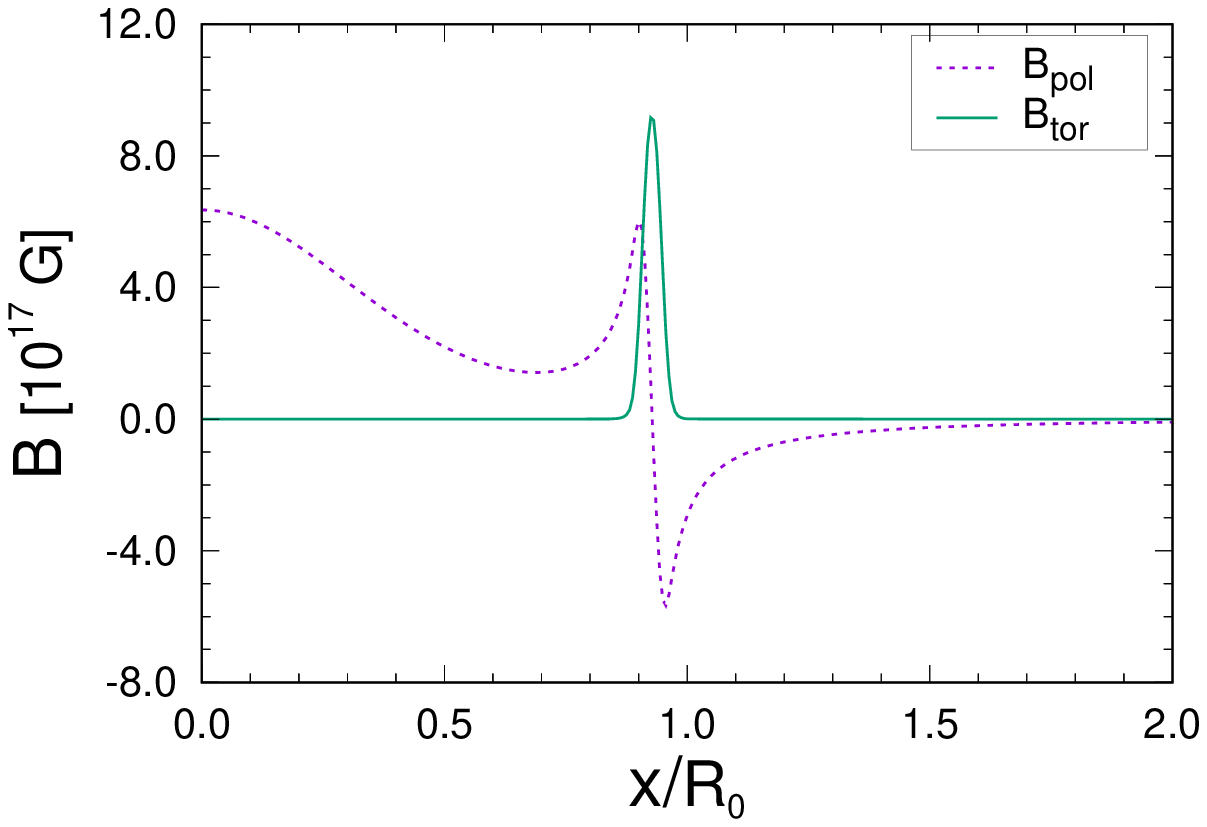}
\\
\includegraphics[height=36mm]{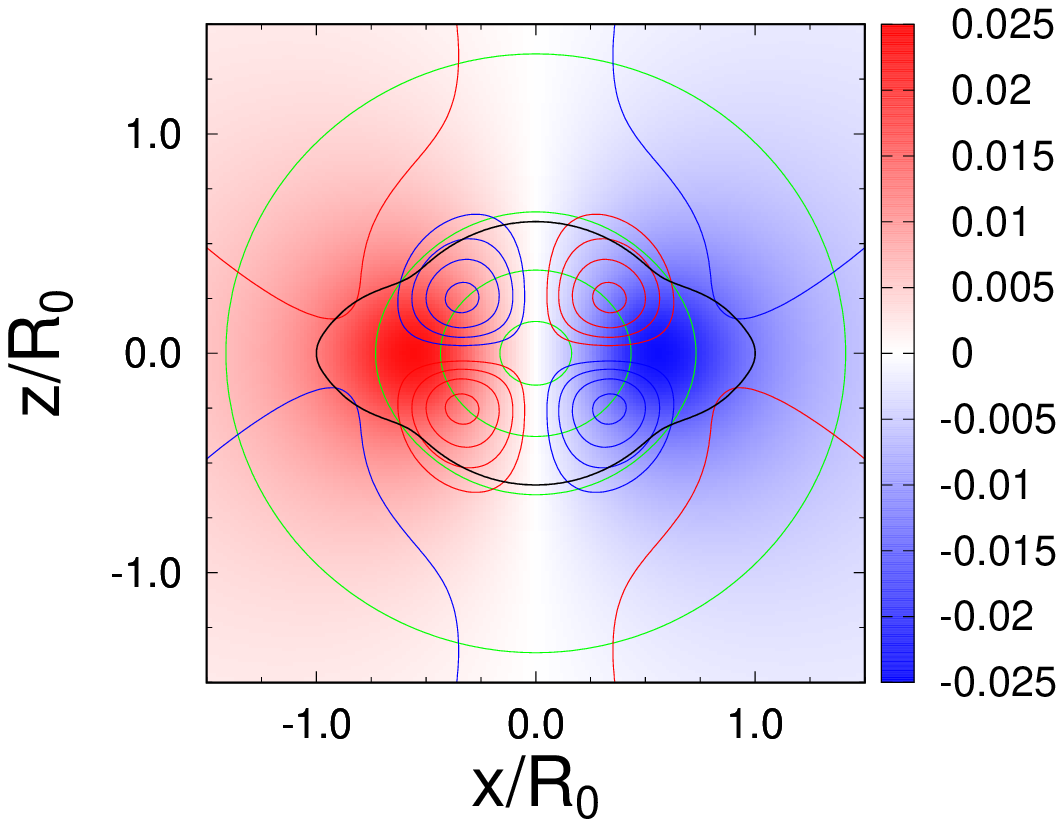}
\includegraphics[height=36mm]{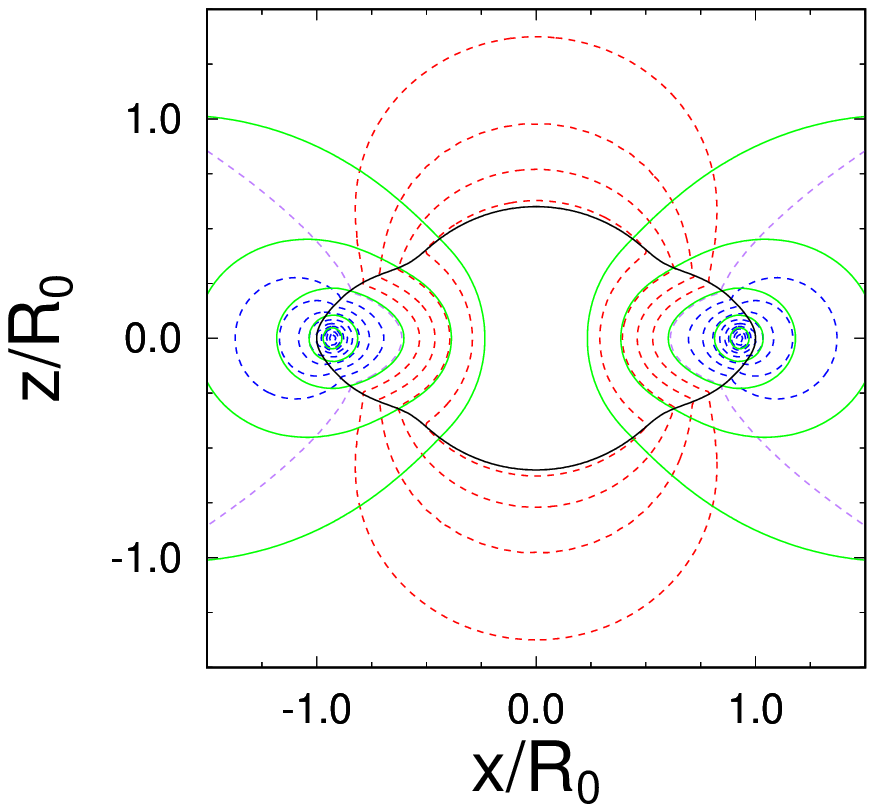}
\includegraphics[height=36mm]{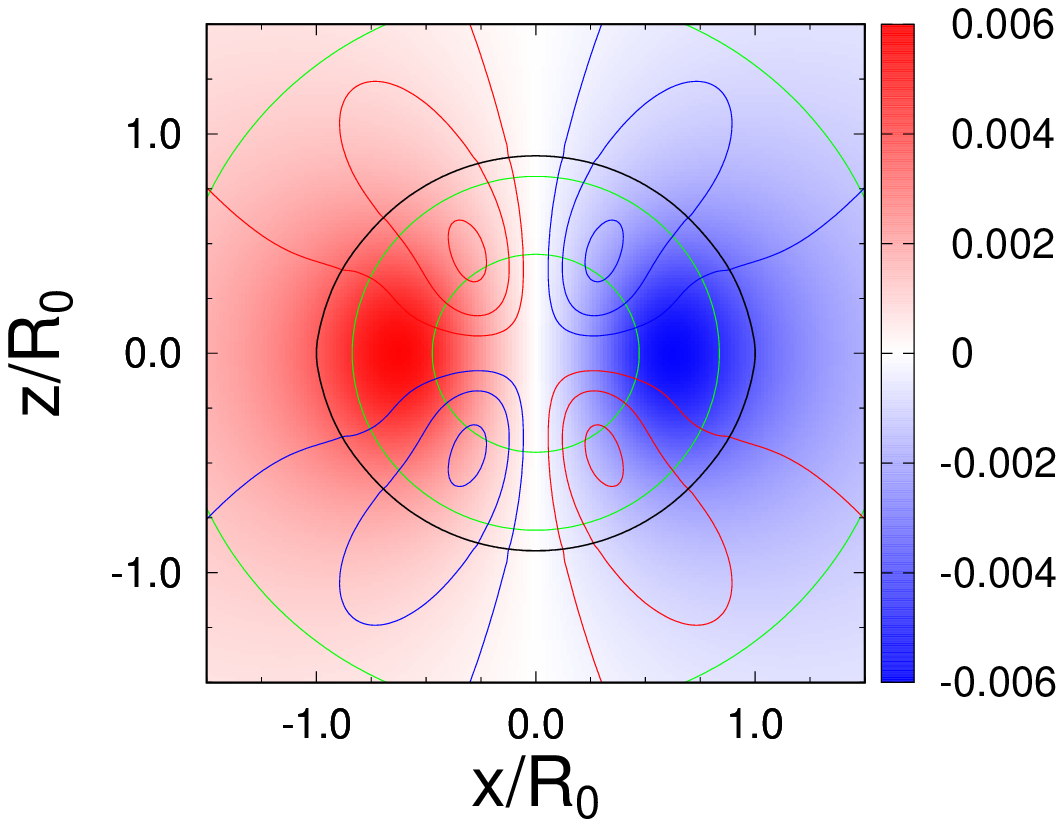}
\includegraphics[height=36mm]{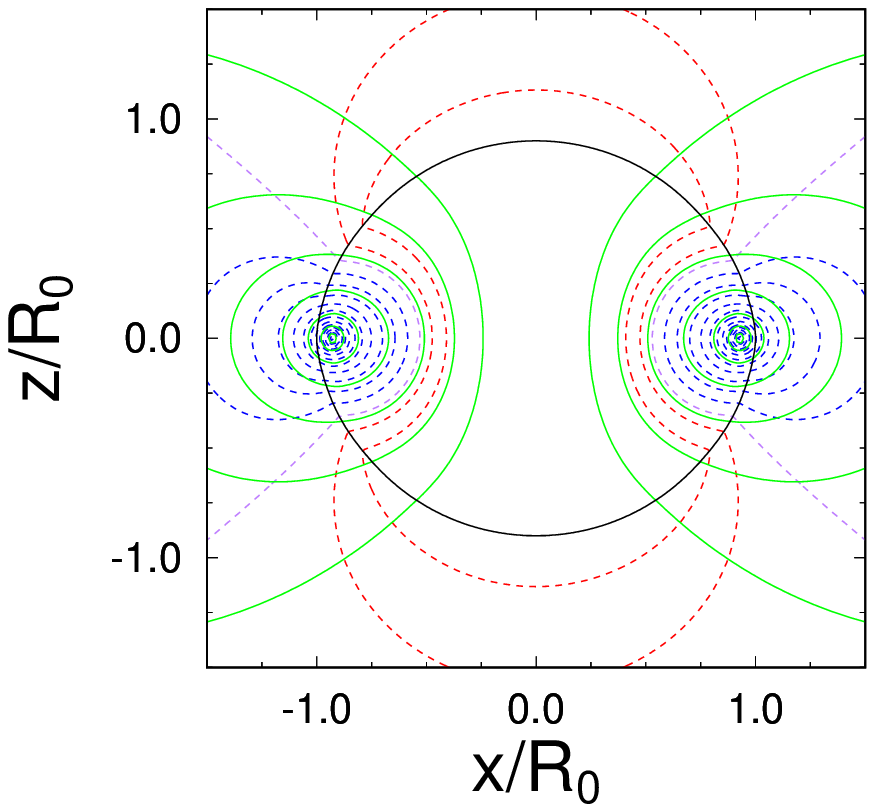}
\caption{
Same as Fig.\ref{fig:EV-MT-UR} but for differentially rotating 
and extremely magnetized compact stars 
associated with an electromagnetic vacuum outside and a magnetotunnel, EV-MT-DR-1 
(rapidly rotating model) and EV-MT-DR-2 (slowly rotating model).  
}  
\label{fig:EV-MT-DR}
\end{center}
\end{figure*}

\begin{figure*}
\begin{center}
\includegraphics[height=42mm]{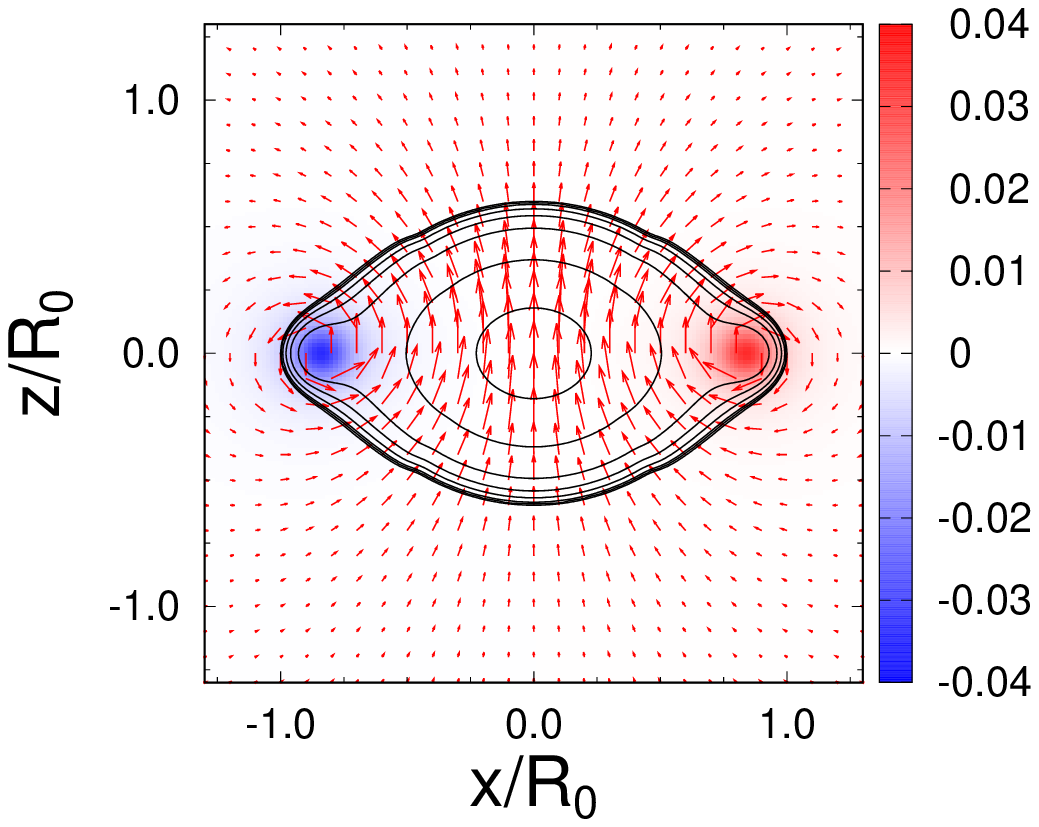}
\includegraphics[height=42mm]{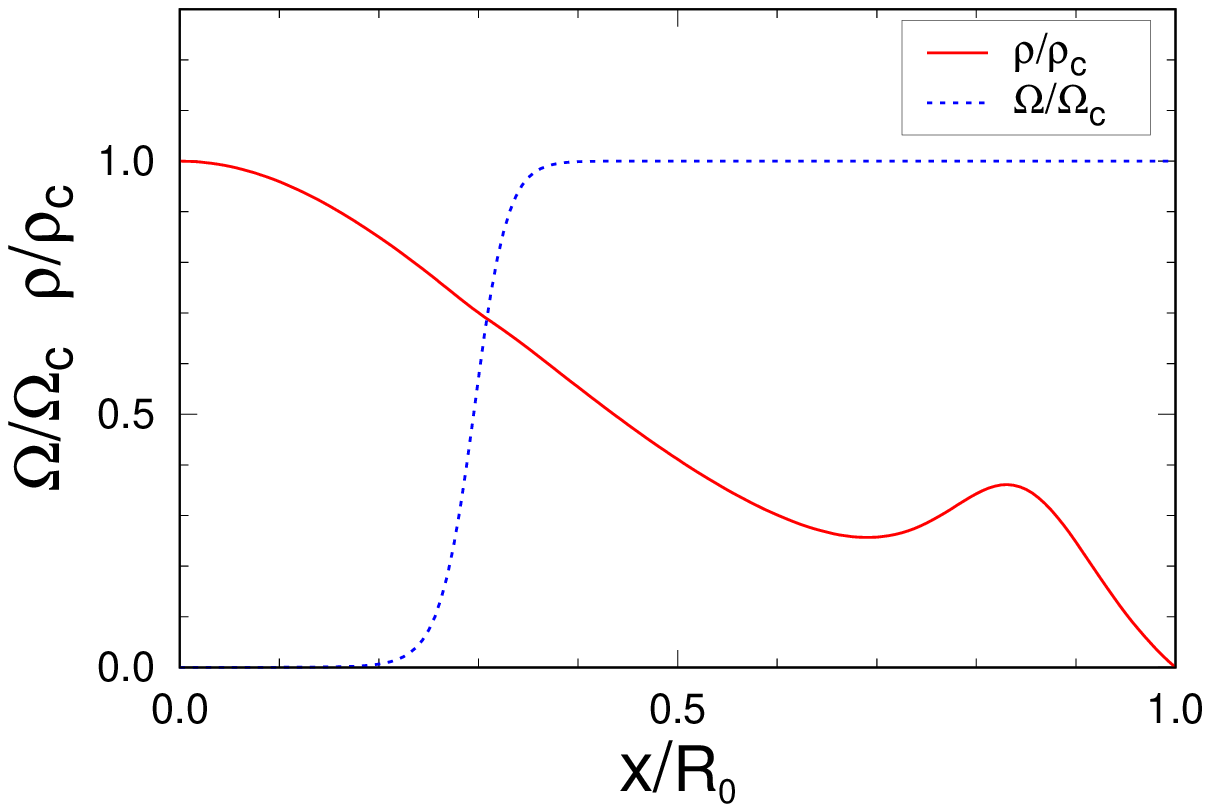}
\includegraphics[height=42mm]{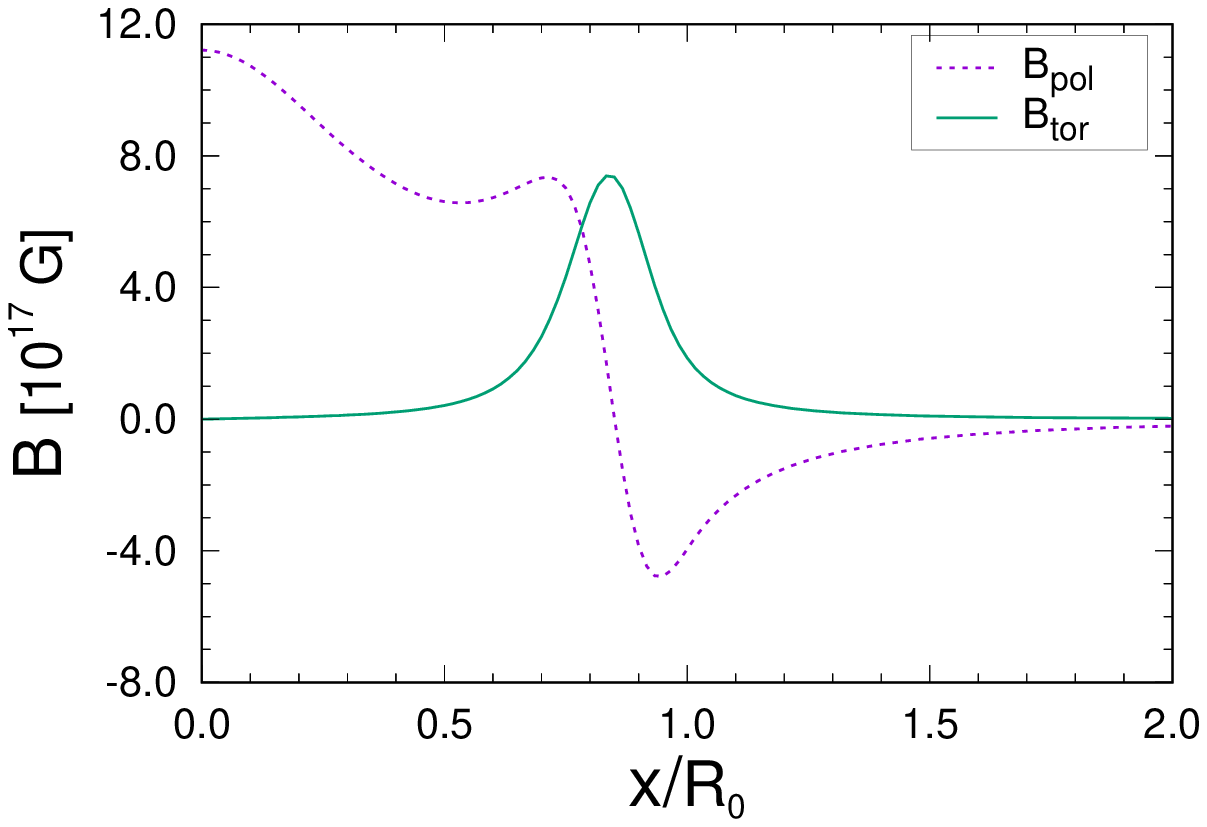}
\\
\includegraphics[height=42mm]{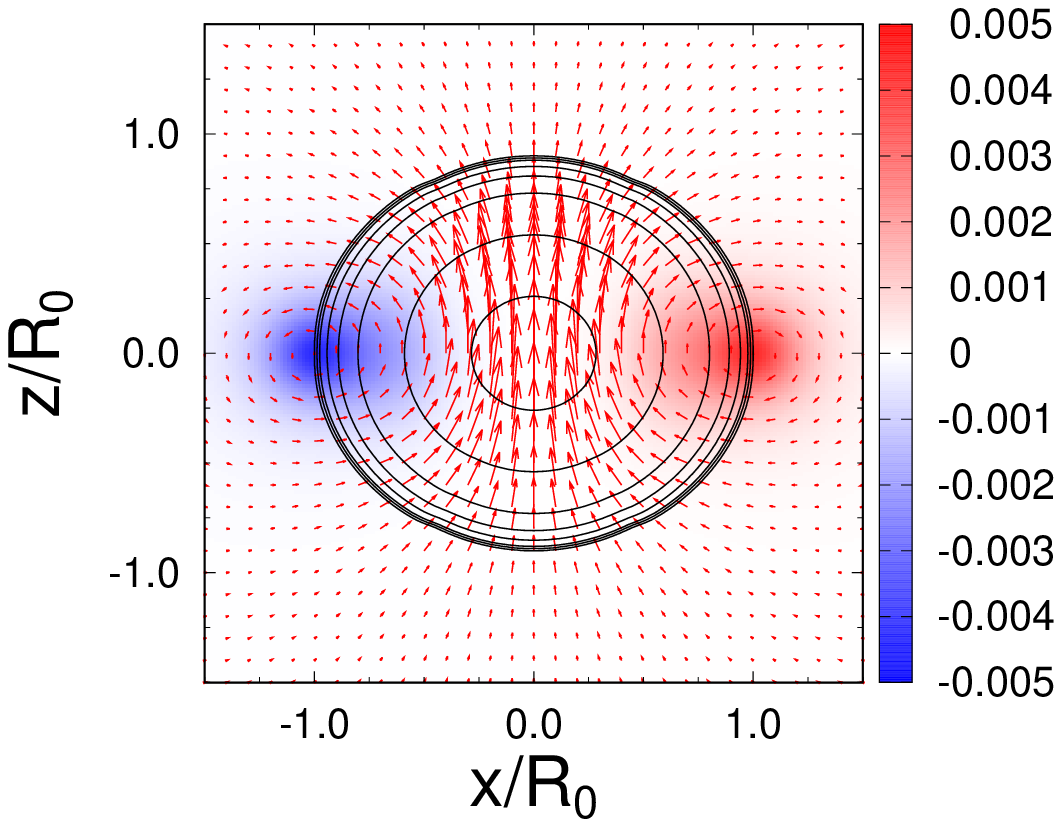}
\includegraphics[height=42mm]{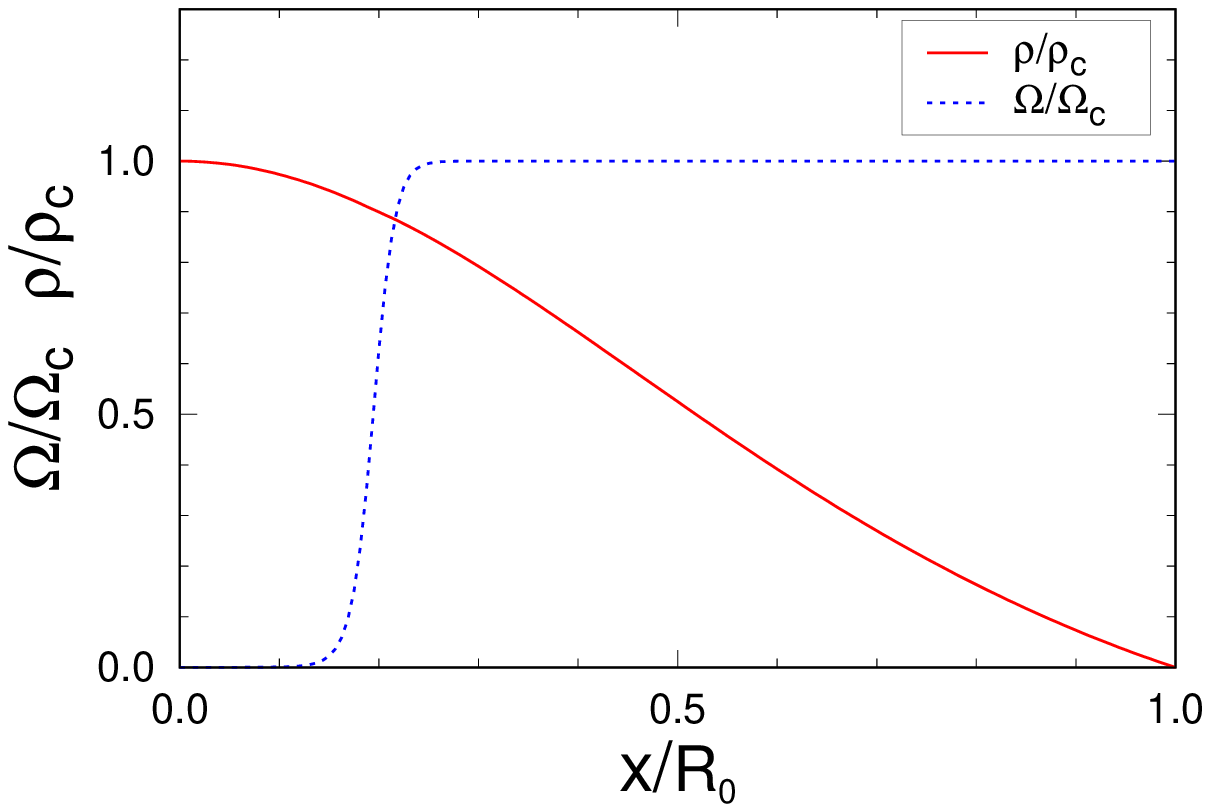}
\includegraphics[height=42mm]{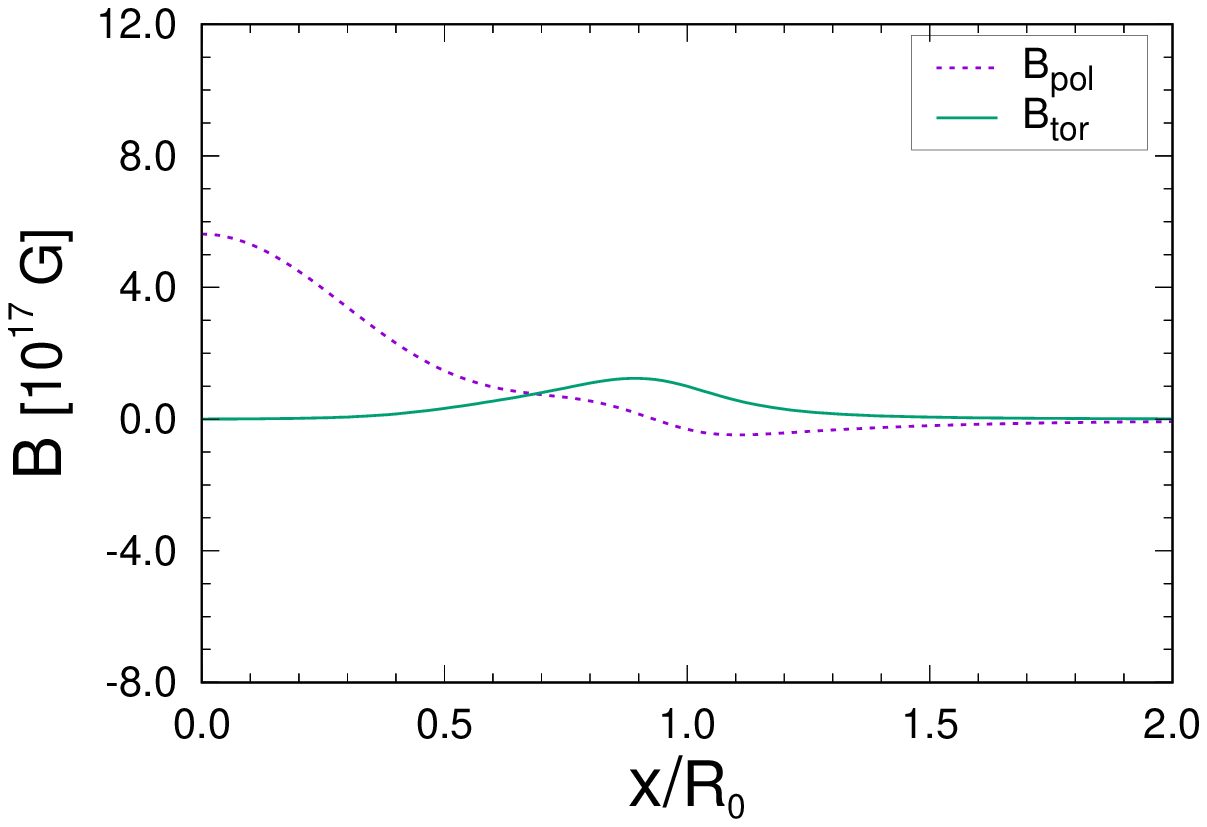}
\\
\includegraphics[height=36mm]{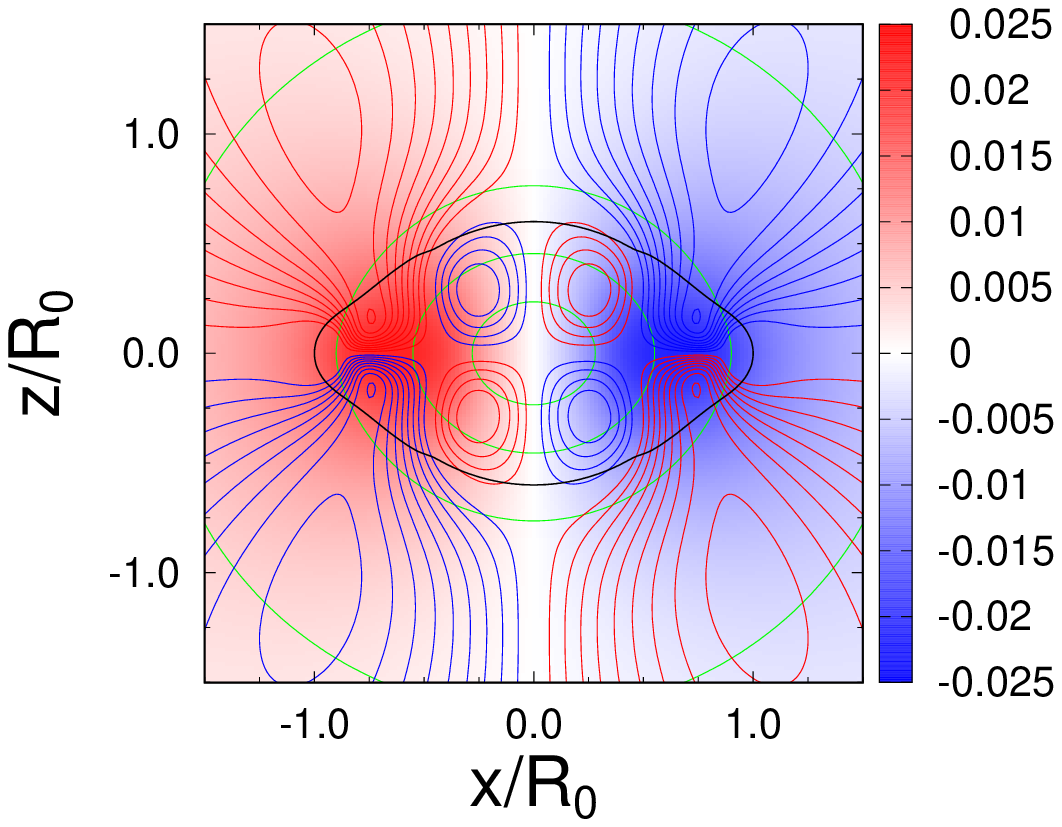}
\includegraphics[height=36mm]{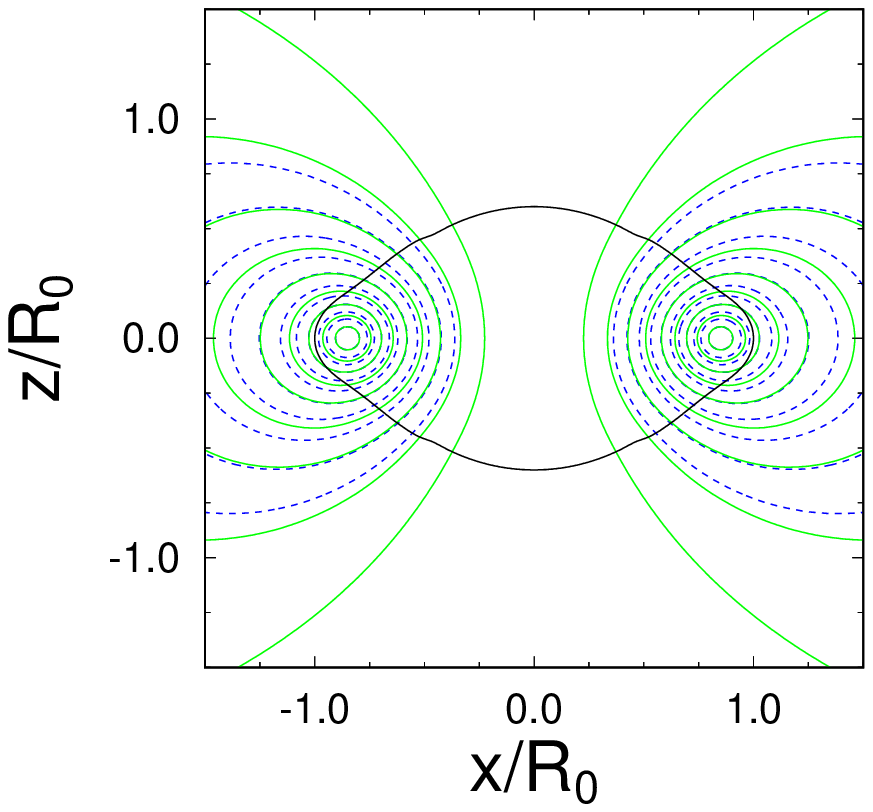}
\includegraphics[height=36mm]{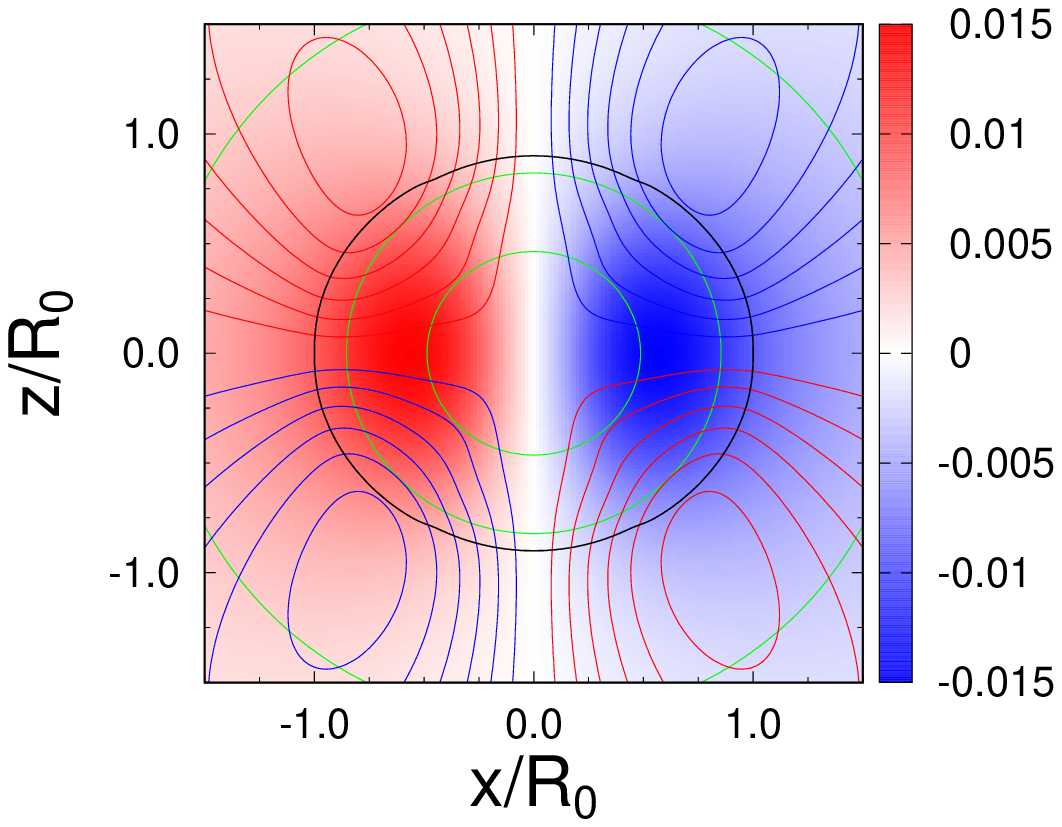}
\includegraphics[height=36mm]{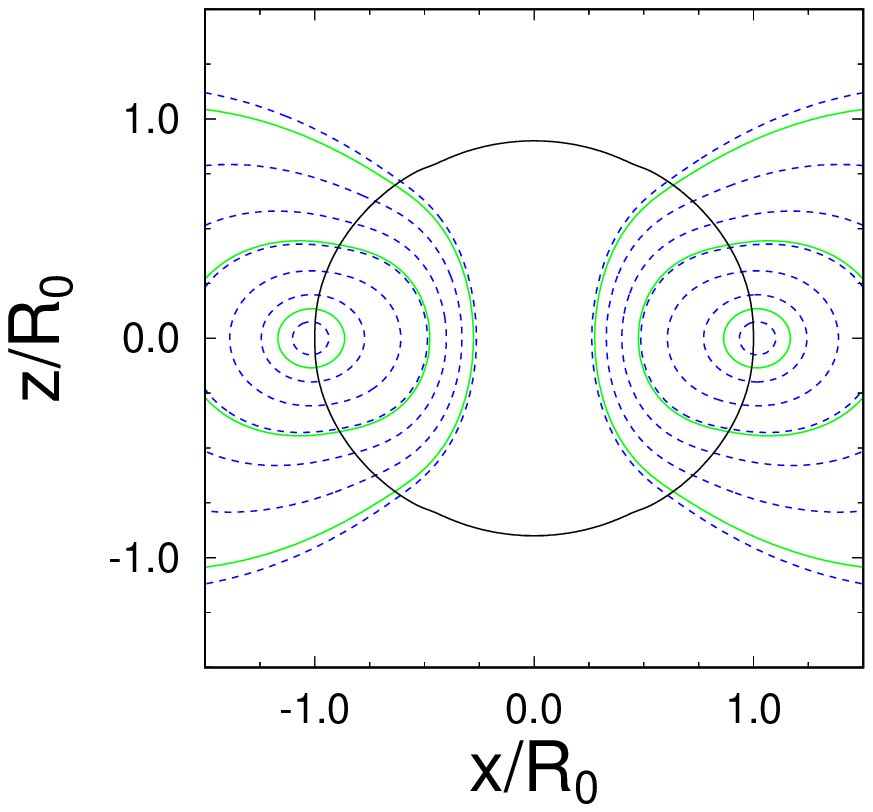}
\caption{
Same as Fig.\ref{fig:EV-MT-UR} but for differentially rotating 
and extremely magnetized compact stars 
associated with a magnetosphere, MS-DR-1 
(supramassive model) and MS-DR-2 (normal mass model), whose 
toroidal magnetic fields are distributed across the stellar support and magnetosphere.  
}  
\label{fig:MS-DR}
\end{center}
\end{figure*}

\begin{table*}[t]
\caption{Physical quantities of solutions presented in 
Figs.~\ref{fig:EV-MT-UR}--\ref{fig:EV-MT-DR}.  
All solutions are calculated with fixing 
the ratio of the maximum values of the pressure to the rest mass density 
$\prhoc = 0.12322$, which corresponds to the rest mass density 
$\rhoc = 1.0717\times 10^{15}\ [\mbox{g/cm}^3]$.  Listed quantities are 
the ratio of the equatorial to polar radii in the coordinate length $R_z/R_0$, 
the equatorial and polar radii in proper length $\bar{R}_0$ and $\bar{R}_z$, 
the angular velocity $\Omega_{\rm c}$ (see, Sec.~\ref{sec:models}), 
the ADM mass $\Madm$, the rest mass $M_0$, 
the angular momentum $J$, 
and a residual of the equality 
of the Komar mass $\MK$ and the ADM mass $\Madm$.  
Units of listed quantities are explained in Table \ref{tab:TOV_solutions}.  
To convert a unit of length from $G=c=\Msol=1$ to [km], multiply 
$G\Msol/c^2=1.477$[km].
Details of the definitions for these quantities are found in Appendix F of Paper I.  
}
\label{tab:MRNS_solutions}
\begin{tabular}{ccccccccc}
\hline
Model &  $R_z/R_0$ &
$\bar{R}_0$ &  $\bar{R}_z/\bar{R}_0$ 
& $\Omegac$ & $\Madm$ & $M_0$ & 
$J/\Madm^2$ & 
$|1-\MK/\Madm|$ 
\\
\hline
EV-MT-UR-1 & 0.6 & 12.496 & 0.62293 & $2.9503\times 10^{-2}$ & 1.4262 & 1.5356 & 0.61783 & $4.3864\times 10^{-5}$ \\
EV-MT-UR-2 & 0.9 & 9.5086 & 0.90536 & $9.6580\times 10^{-3}$ & 1.2097 & 1.2986 & 0.17625 & $1.0482\times 10^{-6}$ \\
MS-MT-DR-1 & 0.6 & 12.601 & 0.62341 & $2.8048\times 10^{-2}$ & 1.4323 & 1.5460 & 0.51691 & $3.9883\times 10^{-3}$ \\
MS-MT-DR-2 & 0.9 & 9.5293 & 0.90550 & $8.4512\times 10^{-3}$ & 1.2066 & 1.2954 & 0.12268 & $2.7386\times 10^{-4}$ \\
EV-MT-DR-1 & 0.6 & 12.972 & 0.62549 & $2.6918\times 10^{-2}$ & 1.3874 & 1.4978 & 0.42490 & $5.8742\times 10^{-3}$ \\
EV-MT-DR-2 & 0.9 & 9.5423 & 0.90571 & $8.2012\times 10^{-3}$ & 1.2054 & 1.2940 & 0.10204 & $3.4697\times 10^{-4}$ \\
MS-DR-1    & 0.6 & 12.301 & 0.62111 & $1.6336\times 10^{-2}$ & 1.5327 & 1.6301 & 0.36932 & $1.1060\times 10^{-3}$ \\
MS-DR-2    & 0.9 & 9.4817 & 0.90481 & $1.4901\times 10^{-2}$ & 1.2189 & 1.3100 & 0.26448 & $1.0416\times 10^{-4}$ \\
\hline
\end{tabular}
\end{table*}

\begin{table*}
\caption{Continued from Table \ref{tab:MRNS_solutions}, 
listed quantities are the maximum values 
of poloidal and toroidal magnetic fields, $\Bpolmax$ and $\Btormax$, 
the ratios of poloidal and toroidal magnetic field energies, 
${\cal M}_{\rm pol}$ and ${\cal M}_{\rm tor}$, and electric field energy, 
${\cal M}_{\rm ele}$, to the total electromagnetic field energy ${\cal M}$, 
the ratios of the kinetic, internal, and electromagnetic field energies 
to the gravitational energy, ${\cal T}/|{\cal W}|$, $\Pi/|{\cal W}|$, 
and ${\cal M}/|{\cal W}|$, respectively, and the virial constant 
$I_{\rm vir}$, and the electric charge contribution from the volume integral 
of the star $Q_M$.  
Details of the definitions are found in Appendix F in Paper I.  
The maximums of magnetic field components $\Bpolmax$ and $\Btormax$ are 
defined by those of spatial Faraday tensor $F_{ab}$ in Cartesian coordinates, 
$\Bpol:=F_{xy}$ and $\Btor:=-F_{xz}$.  
}
\label{tab:MRNS_solutions_EMF}
\begin{tabular}{ccccccccccccc}
\hline
$\Bpolmax$[G] & $\Btormax$[G] & 
${\cal M}_{\rm pol}/{\cal M}$ & ${\cal M}_{\rm tor}/{\cal M}$ &
${\cal M}_{\rm ele}/{\cal M}$ & ${\cal T}/|{\cal W}|$ & $\Pi/|{\cal W}|$ & 
${\cal M}/|{\cal W}|$ & $I_{\rm vir}/|{\cal W}|$ & $Q_M$ 
\\
\hline
$6.1604\times 10^{17}$ & $5.9686\times 10^{17}$ & 0.92731 & 0.036274 & 0.036413  &  0.061677 & 0.28602 & 0.019016 & $4.2789\times 10^{-4}$ & $5.5363\times 10^{-2}$ \\
$6.3763\times 10^{17}$ & $8.4596\times 10^{17}$ & 0.95783 & 0.039465 & 0.0027073 & 0.0052941 & 0.32319 & 0.020126 & $2.9462\times 10^{-4}$ & $9.2879\times 10^{-3}$ \\
$6.0341\times 10^{17}$ & $5.8367\times 10^{17}$ & 0.93139 & 0.031169 & 0.037446  &  0.052395 & 0.28570 & 0.019147 & $1.8954\times 10^{-2}$ & $5.3731\times 10^{-2}$ \\
$6.3494\times 10^{17}$ & $8.6731\times 10^{17}$ & 0.96193 & 0.035946 & 0.0021234 & 0.0033736 & 0.32376 & 0.020848 & $1.1172\times 10^{-3}$ & $7.9114\times 10^{-3}$ \\
$5.4290\times 10^{17}$ & $5.3893\times 10^{17}$ & 0.93710 & 0.035454 & 0.027449  &  0.038567 & 0.29213 & 0.017490 & $2.8983\times 10^{-2}$ & $5.0486\times 10^{-2}$ \\
$6.3568\times 10^{17}$ & $9.1621\times 10^{17}$ & 0.96044 & 0.037997 & 0.0015662 & 0.0025140 & 0.32399 & 0.021508 & $1.4947\times 10^{-3}$ & $7.5440\times 10^{-3}$ \\
$1.1258\times 10^{18}$ & $7.8558\times 10^{17}$ & 0.91411 & 0.070139 & 0.015746  &  0.019785 & 0.26756 & 0.15293   & $4.8131\times 10^{-3}$ & $8.9732\times 10^{-2}$ \\
$5.3348\times 10^{17}$ & $9.8917\times 10^{16}$ & 0.92365 & 0.070049 & 0.0062992 &  0.012604 & 0.32182 & 0.0090924 & $2.4980\times 10^{-4}$ & $2.7732\times 10^{-3}$ \\
\hline
\end{tabular}
\end{table*}

\section{Results}
\label{sec:Results}

From the formulation and numerical method described in 
Sec.~\ref{sec:Formulation}, compact star solutions associated 
with extremely strong electromagnetic fields are obtained.  
Overall configurations of the magnetic fields generated 
from this formulation are typically the strong poloidal (dipole like) 
magnetic fields extending from the core of the star to the outside, 
and the poloidal and toroidal magnetic fields 
concentrated in a toroidal region near the equatorial surface.  
In Paper I, we have observed that the latter very strong 
mixed magnetic fields expel the matter near the equatorial surface.  
As mentioned above, we introduce the force-free magnetic 
fields when the matter is totally expelled that the mass density 
becomes negligible in the toroidal region, so 
the force free magnetotunnel is formed inside of the compact star.  

Combining the electromagnetic vacuum or magnetosphere, and 
the uniform or differential rotations, we present four types of such 
extremely magnetized solutions in this article where three of them 
are associated with the magnetotunnel: 
\begin{itemize}
\setlength{\itemsep}{-1pt}
\item
EV-MT-UR type -- those associated with the electromagnetic vacuum outside 
of the star, magnetotunnel, and uniform rotation.  
\item
MS-MT-DR type -- those associated with the magnetosphere, magnetotunnel, and 
differential rotation.  
\item
EV-MT-DR type -- those associated with the electromagnetic vacuum outside, 
magnetotunnel, and differential rotation.  
\item
MS-DR type -- those associated with the magnetosphere, and differential rotation
whose toroidal magnetic field is distributed across the star and the magnetosphere.
\end{itemize}

\subsection{Overall feature of solutions with magnetotunnel}
\label{sec:MT}

\subsubsection{Uniformly rotating models with electromagnetic vacuum outside}
\label{sec:EV-MT-UR}

In Fig.~\ref{fig:EV-MT-UR}, EV-MT-UR models with rapid and slow rotations 
are shown.  These are straightforward extensions of solutions presented in 
Paper I, in particular changing the parameters $\{\Lambda_0, \Lambda_{\phi 0}\}$ 
of the model P1 systematically to achieve stronger electromagnetic fields.  
The boundaries of the force-free magnetotunnel, where the toroidal regions 
where the matter is totally expelled by the magnetic 
fields, are indicated with green circles in the left panel of the first and 
the second rows for rapidly and slowly rotating models, respectively.  
The expelled regions can be also seen clearly in the rest mass density profile 
$\rho/\rho_c$ (red curves) along the equatorial radius $x/R_0$ plotted in 
the middle panels of the first and the second row.  

The right panel of the first and the second rows in Fig.~\ref{fig:EV-MT-UR} 
are the plots of the poloidal, $B_{\rm pol} = F_{xy}$, and 
the toroidal, $B_{\rm tor} = -F_{xz}$, components of magnetic fields along 
the equatorial radius $x/R_0$.  
The maximum of $B_{\rm pol}$ is at the center of the star, while that 
of $B_{\rm tor}$ is near the equatorial surface.  It can be seen that 
the poloidal component $B_{\rm pol}$ takes a large value also around 
this toroidal region near the equatorial surface.\footnote{This structure 
is referred to as the twisted torus magnetic fields.}  
It is also noticeable that, for the slowly rotating EV-MT-UR-2 model, 
the maximum value of $B_{\rm tor}$ is greater than that of $B_{\rm pol}$.  

In the second and fourth panels in the third row, the contours of the 
components of electromagnetic 1-form $A_t$ and $A_\phi$ are drawn.  
Because $A_t$ is assumed to be a function of $A_\phi$ in the ideal MHD 
region of the stellar support satisfying the relation (\ref{eq:MHDfnc_At}), 
their contours are homologous there, while they are not in the 
electromagnetic vacuum outside of the star.

\subsubsection{Differentially rotating models with magnetosphere}
\label{sec:MS-MT-DR}

In Fig.~\ref{fig:MS-MT-DR}, MS-MT-DR models with rapid and slow rotations 
are shown.  The same as the EV-MT-UR model in Fig.~\ref{fig:EV-MT-UR}, 
the strong concentration of the magnetic fields 
expels the matter near the equatorial surface and form the magnetotunnel, 
which can be seen in the corresponding panels in Fig.~\ref{fig:MS-MT-DR}.  

Differences between MS-MT-DR model and the previous EV-MT-UR model 
are the magnetosphere outside of the star instead of electromagnetic vacuum, 
and the differential rotation instead of the uniform rotation.  
As discussed in Sec.~\ref{sec:Formulation}, arbitrary functions 
in the force-free magnetosphere/magnetotunnel region are chosen to be 
the same as those in ideal MHD region as Eq.~(\ref{eq:iMHD_FF}) that 
the electromagnetic potentials are smoothly connected across the boundaries 
of these regions.  This can be seen in 
the second and the fourth panels of the third row in Fig.~\ref{fig:MS-MT-DR} 
for the contours of $A_t$ and $A_\phi$, which are homologous not only 
on the stellar support (ideal MHD region) but also in the force-free region 
outside of the star (as well as in the magnetotunnel region).  

We also introduce the differential rotation as prescribed in 
Eq.~(\ref{eq:MHDfnc_Omega_FF}) in this model.  As shown in 
the middle panels of the first and the second row of 
Fig.~\ref{fig:MS-MT-DR}, the profiles of $\Omega/\Omega_c$ are 
\emph{increasing} along the equatorial radius $x/R_0$.  
These are rather uncommon profiles for differential rotation laws, 
that is, in most of numerical computations of relativistic rotating 
stars, differential rotation laws with decreasing $\Omega$ along 
$x/R_0$ have been considered \cite{RNSdiff}, except for a few works 
\cite{RNSdiff2}.  
There are two motivations for us to choose this new differential 
rotation law (\ref{eq:MHDfnc_Omega_FF}).  

In computing solutions with force-free magnetosphere outside, 
we found that the $t$-component of electromagnetic 1-form $A_t$
diverges asymptotically if we assume uniform rotation.  
This may be understood that in our assumptions, namely the stationary 
and axisymmetric force-free magnetosphere, the magnetic field lines 
attached to the stellar surface have to rotate with the same $\Omega$ 
even if they are extended to the region outside of the light cylinder.  
In terms of our formulation, it appears that 
the current $j^t$ coupled with the $\phi$-component $j^\phi$ as in 
Eq.~(\ref{eq:current_jphi_FF}) doesn't fall-off asymptotically fast 
enough to have a regular asymptotic behavior in $A_t$.  
Since the magnetic field lines extended towards the region of 
larger $r$ are attached to the surface of the star closer to the axis of 
rotation ($z$-axis), setting $A_t' = \Omega(A_\phi)$ to vanish near 
the rotation axis decouples the $j^\phi$ and $j^t$ in the large $r$ region.  
This is the first reason that we choose the differential 
rotation law (\ref{eq:MHDfnc_Omega_FF}), which is non-rotating near 
the axis of symmetry.
Also for this reason, it seems it is not possible to obtain uniformly 
rotating relativistic solutions with the force-free magnetosphere using 
our formulation.  In the literature, only 
the non-rotating solutions are calculated for such strongly magnetized 
relativistic stars associated with the force-free magnetosphere in 
general relativity \cite{Pili:2014zna}.  
The second reason for this differential rotation law is motivated by 
the results of simulations \cite{Tsokaros:2021pkh} and 
a semi analytic argument \cite{Shapiro:2000zh}
that such rotation profiles may, although transiently, 
appear during the evolution of highly magnetized rotating stars.  

Because of the differential rotation, the rotation period of the field 
lines in the magnetosphere differs with latitude.  The fastest ones are 
those attached near the equatorial surface.  The smallest cylindrical radius 
of the light cylinder becomes around 
$\varpi=r \sin\theta \sim 2\pi/\Omega_c \sim 24R_0$ $(100R_0)$ for 
the rapidly (slowly) rotating model MS-MT-DR-1 (2, respectively), 
so the rotating field lines do not reach to the light cylinder in our models.

\subsubsection{Differentially rotating models with electromagnetic vacuum outside}
\label{sec:EV-MT-DR}

In Fig.~\ref{fig:EV-MT-DR}, EV-MT-DR models with rapid and slow rotations 
are shown.  This is to demonstrate that it is possible to calculate 
solutions combining the electromagnetic vacuum region outside and 
the differential rotation.  Because of the magnetic vacuum, the field lines 
outside are not dragged around.  Although in principle, one can freely specify 
the differential rotation law for these models, we only modify the values of 
parameters slightly from MS-MT-DR models.  
It appears that interior magnetic fields of EV-MT-DR models are similar to 
those of MS-MT-DR models rather than EV-MT-UR models.

\subsection{
Solutions with toroidal fields distributed across the star and 
the magnetosphere
}
\label{sec:MS-DR}

Because of our previous choices of parameter, in particular $A_\phi^0$ 
for the function (\ref{eq:MHDfnc_dXi}) (see Table \ref{tab:para_functions}), 
the function varies when the potential $A_\phi$ becomes 
larger than its value at the equatorial surface, 
$A_\phi > A_\phi^0 = A_{\phi, \rm S}^{\rm max}$.  
Because of this, the toroidal component of the magnetic field is 
confined interior of the compact star.  This choice was necessary for the 
EV models since the toroidal magnetic field can not exist in the vacuum region.  
Also because of this, the extremely strong magnetic fields 
develop near the equatorial surface, which is strong enough to expel 
the matter as shown in Sec.~\ref{sec:MT}.  

For the MS models, however, the toroidal 
component of the magnetic fields is allowed to exist in the 
region of force-free magnetosphere outside of the star.  
In Fig.~\ref{fig:MS-DR}, we successfully computed such strongly magnetized 
solutions whose toroidal component of magnetic fields is distributed across 
the star and the magnetosphere.  For these solutions with the magnetosphere, 
the parameter $A_\phi^0$ is chosen to be $A_\phi^0 = 0.3 A_{\phi, \rm S}^{\rm max}$, 
and $A_\phi^1$ is also modified as $A_\phi^1 = 1.7 A_{\phi}^{\rm max}$ 
(see, Table \ref{tab:para_functions}).  

As seen in Fig.~\ref{fig:MS-DR}, the peak of toroidal component near 
the equatorial surface becomes broader and less concentrated 
compared with the other magnetotunnel models in previous 
Sec.\ref{sec:MT}.  For these MS-DR models, we couldn't find a solution 
with magnetotunnel in a parameter region we searched solutions.  
As shown in Fig.~\ref{fig:MS-DR}, for the largely deformed model 
MS-DR-1, the matter is expelled in a wider region, but not totally.  
For the less deformed model MS-DR-2, on the other hand, we couldn't 
compute a solution with the same parameter, but obtained a solution 
with smaller and broader peak of the toroidal component.  
The maximum of the toroidal component of this model is close to 
the stellar surface.

\subsection{Physical quantities of solutions}

In Tables \ref{tab:MRNS_solutions} and \ref{tab:MRNS_solutions_EMF}, 
physical quantities of solutions presented in 
Figs.~\ref{fig:EV-MT-UR}-\ref{fig:MS-DR} are listed.  
For all models (for both of rapidly and slowly rotating cases), we choose 
the same central (maximum) rest mass density $\rho_c$.  
As shown in the tables, for the solutions with the magnetotunnel, 
the rest mass $M_0$ are around $1.5\Msol$ and $1.3\Msol$ for the rapidly 
and slowly rotating models, respectively.  
Hence, in our unit (choice of the polytropic constant $K$), 
corresponding non-rotating and non-magnetized solutions, that is spherically 
symmetric TOV solutions, with the same rest mass have the compactness around 
$\compa\sim0.2$ and $\compa\sim 0.15$, respectively.  
Therefore, these solutions are mildly compact models.  
On the other hand, the MS-DR-1 is a supramassive model associated 
with the strongest electromagnetic fields among other models.

From the virial relation with an equality $\Madm=\MK$ \cite{virial}, 
\beq
I_{\rm vir}\,:=\,2{\cal T}+3\Pi+{\cal M}+{\cal W}=0, 
\eeq
one can roughly understand the contribution of the kinetic term 
${\cal T}$ and electromagnetic term ${\cal M}$ to the deformation of 
the compact stars.  For the solutions with the magnetotunnel, 
$2 {\cal T}$ are around 4-6 times of ${\cal M}$ for the rapidly rotating models, 
while for the slowly rotating models ${\cal M}$ is dominating about 
2-3.5 times over $2 {\cal T}$.  
The ratios of $(2{\cal T}+{\cal M})/3\Pi$ which roughly 
measure contribution of non-spherical deformation to the equilibriums 
are about 3\% for the slowly rotating models, and 15\% for the rapidly 
rotating models. 
For the MS-DR-1 model, on the other hand, ${\cal M}$ is dominating 
over $2 {\cal T}$, where ${\cal M}$ is about 4 times larger than $2 {\cal T}$.  
The value of ${\cal M}/|\cal W|$ is about 7 times larger than the other models.

It is also commented that, for the models with the magnetotunnel, 
the maximum values of the toroidal components of 
magnetic fields are comparable or even larger than that of poloidal magnetic 
fields, overall integrals of toroidal fields ${\cal M}_{\rm tor}$ are 
only 3.5-4\% of those of poloidal fields ${\cal M}_{\rm pol}$.
The integrals of electric part ${\cal M}_{\rm ele}$ is about the 
same as ${\cal M}_{\rm tor}$ for the rapidly rotating models, but 
it is less than 10\% of ${\cal M}_{\rm tor}$ for the slowly rotating 
models.  This seems to be reasonable considering that the higher 
multipole contributions are less dominating in the slowly rotating models.  
For the MS-DR models, 
the toroidal component of magnetic field is distributed in broader region, 
and hence the fraction of ${\cal M}_{\rm tor}/{\cal M}$ is 
about twice of the other models, although it is still more than an order 
smaller than the contribution from the poloidal component ${\cal M}_{\rm pol}/{\cal M}$.

\section{Discussion}
\label{sec:Discussion}

Results of simulations by Braithwaite and co-workers
\cite{Braithwaite} suggest that stable equilibriums of 
strongly magnetized stars may be achieved when the energies 
of the poloidal and the toroidal components of magnetic fields 
become comparable \cite{Braithwaite}.  One of motivations to 
investigate the solutions of such mixed poloidal and toroidal 
magnetic fields presented in this paper is to obtain 
such stable models of magnetized compact stars.  
However, so far, the energy carried by the toroidal field is 
far smaller than that of the poloidal field in our models.  
Recently we have performed numerical simulations of such 
extremely magnetized compact stars starting from the initial 
data calculated in Paper I (and with varied parameters) which 
are close to the EV-MT-UR models but with $\Btormax$ around 30-40\% 
smaller \cite{Tsokaros:2021pkh}.  We found the kink instability 
develop and destroy the axisymmetry of the solutions, although 
in a certain case the instability develop slower than the alfv\'en 
time.  Also found was the magnetorotational effect carries away 
the angular momentum of the stellar core, hence the rotation 
of the core slows down and a differential rotation develops.  
It is totally unclear, but is interesting to investigate, how 
the stronger toroidal magnetic field, and/or a differential rotation 
as in the present models modify the evolutions of such compact stars.

The Magnetic field strength of the solutions presented in this paper 
may be too strong for astrophysically realistic compact objects.  
From a theoretical stand point, however, it is of interest to investigate 
the extreme cases where the electromagnetic fields affect the stellar 
equilibrium or even become a source of gravity.  
In the above solutions, it is observed that the magnetic fields locally 
dominate over the hydrostatic equilibriums, but the metric is affected 
only slightly.  As seen in the contours of metric components in 
Figs.~\ref{fig:EV-MT-UR}-\ref{fig:MS-DR}, in the toroidal region 
near the equatorial surface where the strong magnetic fields are 
concentrated, the contour for $\psi$ and the density map of $\tbeta_y$ 
appears to be unaffected, while some structure is observed in the 
contours of $h_{xz}$ in this region.  
Hence, the limit of the strength of magnetic field is not reached 
in a sense that it is not a dominant source of gravity.  
Since our numerical method solves the full 
set of Einstein's and Maxwell's equations for equilibrium or quasi-equilibrium 
initial data, we expect that even more extreme magnetic fields may be obtained, 
including an extremely strong magnetosphere surrounding a black hole.  
Such studies may be one of future extensions of the present work.

\acknowledgments
This work was supported by 
JSPS Grant-in-Aid for Scientific Research(C) 22K03636, 18K03624, 21K03556 
18K03606, 17K05447, 20H04728, NSF Grant PHY-1662211, NASA Grant 80NSSC17K0070, 
and the Marie Sklodowska-Curie grant agreement No.753115.  
AT acknowledges support from the National Center
for Supercomputing Applications (NCSA) at the University of Illinois at
Urbana-Champaign through the NCSA Fellows program.
%


\end{document}